\pdfoutput=1
\newcommand*{\ATLASLATEXPATH}{latex/}

\documentclass[UKenglish,cernpreprint,txfonts,texlive=2016]{\ATLASLATEXPATH atlasdoc}

\DeclareOldFontCommand{\rm}{\normalfont\rmfamily}{\mathrm}
\DeclareOldFontCommand{\bf}{\normalfont\bfseries}{\mathbf}

\usepackage{\ATLASLATEXPATH atlaspackage}
\usepackage{\ATLASLATEXPATH atlasbiblatex}

\usepackage{\ATLASLATEXPATH atlascontribute}

\usepackage{\ATLASLATEXPATH atlasphysics}

\graphicspath{{logos/}{figures/}}

\usepackage{Paper-defs}

\usepackage{nicefrac}
\usepackage{multirow}
\usepackage[colorinlistoftodos,prependcaption]{todonotes}
\usepackage{verbatim}

\def\Wjets{\ensuremath{W\textrm{+jets}}\xspace}

\AtlasTitle{
Search for pair production of vector-like top quarks in events with one lepton, 
jets, and missing transverse momentum 
in $\sqrt{s} = 13$\,\TeV\ \textit{pp} collisions with the ATLAS detector}

\author{The ATLAS Collaboration}

\AtlasRefCode{EXOT-2016-15}

\PreprintIdNumber{CERN-EP-2017-075}

\AtlasJournal{JHEP}

\AtlasAbstract{%
  The results of a search for vector-like top quarks 
using events with exactly one lepton, at least four jets, and large missing transverse momentum are reported. 
The search is optimised for pair production of vector-like top quarks in the $Z(\rightarrow \! \! \nu \nu) \, t + X$ decay channel. 
LHC \textit{pp} collision data at a centre-of-mass energy of $\sqrt{s}=13$\,\TeV\ recorded by the ATLAS detector in 2015 and 2016 are used, corresponding to an integrated luminosity of 36.1\,\ifb. 
No significant excess over the Standard Model expectation is seen and upper limits on the 
production cross-section of a vector-like $T$ quark pair as a function of the $T$ quark mass are derived.
  The observed (expected) 95\% CL lower limits on the $T$ mass are 870\,\GeV (890\,\GeV) for the weak-isospin singlet model, 1.05\,\TeV (1.06\,\TeV) for the weak-isospin doublet model and 1.16\,\TeV (1.17\,\TeV) for the pure $Zt$ decay mode. Limits are also set on the mass as a function of the decay branching ratios, excluding large parts of the parameter space for masses below 1\,\TeV.
}

\hypersetup{pdftitle={ATLAS document},pdfauthor={The ATLAS Collaboration}}

\begin{document}

\maketitle



\section{Introduction}
\label{sec:intro}

A number of theories beyond the Standard Model (SM) of particle physics address the naturalness problem~\cite{Naturalness} and offer mechanisms through which the quadratic divergences, which arise from the radiative corrections to the Higgs boson mass, are resolved. 
A straightforward extension of the SM is the inclusion
of a heavy fourth generation. However, fourth-generation quarks
with SM-like chiral couplings are excluded as they contribute through loops to the couplings of the Higgs boson, 
altering the Higgs boson production cross-sections
to values incompatible with observation~\cite{Djouadi:2012ae,Eberhardt:2012gv}.  
These constraints on chiral quarks can be evaded by vector-like quarks (VLQs)~\cite{1306-0572,1304-0579}, 
hypothetical spin-1/2 coloured particles
whose left-handed and right-handed states have the same electroweak coupling. Vector-like quarks could 
dampen the unnaturally large quadratic corrections to the Higgs boson mass by contributing significantly to loop corrections.
They appear mainly in the ``Little Higgs''~\cite{ArkaniHamed:2001nc,LittleHiggsReview} and ``Composite Higgs''~\cite{CompositeHiggs} 
classes of models.

In this analysis, a vector-like top quark partner ($T$) with a charge of $ Q=+\nicefrac{2}{3}\,|e|$ is searched for. 
When assuming couplings to only the third 
generation of quarks~\cite{Protos}, as favoured by their large masses~\cite{delAguila:1982fs}, 
the three possible decay modes are $T\rightarrow Zt$, $T\rightarrow Ht$, and $T\rightarrow Wb$.
For $T$ quark masses above about 0.8\,\TeV the branching ratios converge to $\BRto{Wb}\simeq50\%$ and $\BRto{Zt}\simeq\BRto{Ht}\simeq25\%$ for 
a weak-isospin singlet model and to $\BRto{Zt}\simeq\BRto{Ht}\simeq50\%$ for a model with a weak-isospin doublet of vector-like quarks.
The doublet prediction is valid for a $(X^{5/3}, T)$ doublet, where the superscript refers to the charge of the particle, as well as for a $(T, B^{-1/3})$ doublet, 
as long as the generalised CKM (Cabibbo-Kobayashi-Maskawa) matrix elements fulfil $|V_{Tb}| \ll |V_{Bt}|$~\cite{Protos,ProtosPage}. In addition to these two models, the case of arbitrary branching ratios of the three possible decay modes is considered.

\begin{figure}[htb]
	\centering
	\includegraphics[width=.5\textwidth]{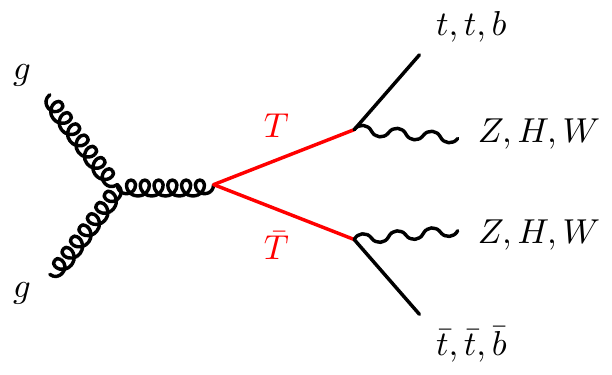}
	\caption{Representative diagram for the production and decay of a vector-like top quark pair.}
	\label{fig:feynman-graph}
\end{figure}

Vector-like top (VLT) quarks could be produced singly or in pairs at the CERN Large Hadron Collider (LHC)~\cite{LHCPaper}. This analysis targets the pair production $pp \rightarrow T\bar{T}$ channel, as shown in Figure~\ref{fig:feynman-graph}. The production cross-section is predicted to be $0.044\pm0.005$\,pb for a $T$ quark mass of 1\,\TeV, 
calculated at next-to-next-to-leading order (NNLO) in quantum chromodynamics (QCD) including next-to-next-to-leading logarithmic (NNLL) soft-gluon resummation 
using \textsc{Top++}~2.0~\cite{Czakon:2011xx,Cacciari:2011hy,Baernreuther:2012ws,Czakon:2012zr,Czakon:2012pz,Czakon:2013goa}.
Previous searches for pair-produced vector-like $T$ quarks by the ATLAS and CMS collaborations at centre-of-mass 
energies of 8\,\TeV~\cite{EXOT-2013-16,EXOT-2013-17,EXOT-2013-18,CMS8TeV} 
and 13\,\TeV~\cite{SUSY-2015-02}
set lower limits on the VLT mass in the range of 550--900\,\GeV, at the 95\% confidence level (CL), depending on the branching ratio considered.

This analysis focuses on the channel $T\bar T \rightarrow Zt + X$, where the $Z$ boson decays into a neutrino pair 
and exactly one charged lepton is produced in either the top quark decay or from the~other~leg~of~the~VLT~pair decay. This channel gives rise to 
a final state with one lepton, multiple jets and a high missing \mbox{transverse} momentum (\met) mainly due to the invisibly decaying $Z$ boson. 
A single-bin signal region (SR) is defined by an event selection that maximises the sensitivity to this particular channel.
Control regions (CRs) and validation regions (VRs) are defined that are enriched in the various background processes. 
They are orthogonal to the SR and orthogonal to each other.
The statistical interpretation is based on a simultaneous fit to the CRs and the SR, in which the background normalisations 
and a possible signal contribution are determined, while taking into account the experimental and theoretical \mbox{systematic uncertainties}.


\section{ATLAS detector and data sample}
\label{sec:detector}
The ATLAS detector~\cite{PERF-2007-01} is a multi-purpose particle detector at the LHC with nearly $4\pi$ coverage around the collision point.\footnote{
	ATLAS uses a right-handed coordinate system with its origin at the nominal interaction point (IP) in the centre of the detector
	and the $z$-axis along the beam pipe. The $x$-axis points from the IP to the centre of the LHC ring, and the $y$-axis points upward. Cylindrical coordinates ($r$,$\phi$) are used in the transverse plane, $\phi$ being the azimuthal angle around the beam pipe.
	The pseudorapidity is defined in terms of the polar angle $\theta$ as $\eta= -\ln\tan(\theta/2)$. Angular distance is measured in units of $\Delta R = \sqrt{(\Delta \eta)^2+(\Delta \phi)^2}$.}
Closest to the beam is the inner detector (ID), which provides charged-particle tracking in the range $|\eta| < 2.5$. During the LHC shutdown between Run 1 and Run 2, a new innermost layer of silicon pixels was added, which improves the track impact parameter resolution and vertex position resolution performance~\cite{ATLAS-TDR-19,ATL-PHYS-PUB-2015-051}. 
 The ID is surrounded by a superconducting solenoid providing a 2\,T axial magnetic field, 
followed by an electromagnetic lead/liquid-argon (LAr) sampling calorimeter and a hadronic (steel/scintillator-tile) calorimeter. 
The endcap and forward regions are instrumented with LAr calorimeters for both the electromagnetic and hadronic energy measurements up to $|\eta|=4.9$.
The outer part of the detector consists of a muon spectrometer (MS) with high-precision tracking chambers for coverage up to $|\eta|=2.7$,
fast detectors for triggering over $|\eta| < 2.4$, and three large superconducting toroidal magnets with eight coils each.
Events are selected by a two-level trigger system consisting of a hardware-based trigger for the first level and a software-based system for the second level~\cite{TRIG-2016-01}.

The data used in this analysis were recorded in 2015 and 2016, corresponding to an integrated luminosity of 36.1\,\ifb\ in total, with an uncertainty of 3.2\%, derived following a methodology similar to that in \mbox{Ref.~\cite{lumi2}}. The data were collected using triggers that select events with high missing transverse momentum, with online thresholds of 70\,\GeV in 2015 and 90\,\GeV to 110\,\GeV in 2016. 
The triggers have an efficiency greater than 99\% for an offline \met requirement of at least 300\,\GeV.

\section{Signal modelling and Monte Carlo simulation}
\label{sec:mc_simulation}
Monte Carlo (MC) simulated events are used for the description of the backgrounds and to model the VLT signals. Signal samples are generated at leading order (LO) with \PROTOS v2.2~\cite{Protos,ProtosPage}, interfaced with \PYTHIA 8.186~\cite{Pythia8} for the parton shower (PS) and hadronisation. 
The samples are produced assuming vector-like $T$ quark couplings according to the weak-isospin singlet model, but 
with equal branching ratios into each of the three decay modes ($Zt$, $Ht$, $Wb$).
Desired branching ratios are obtained by reweighting the samples based on generator information.
To test 
potential 
kinematic biases from the assumed couplings,
a comparison to a sample in which the $T$ quark is in a weak-isospin doublet was carried out.
For a $T$ quark mass of 950\,\GeV a relative acceptance increase of 10\% is observed in the doublet case.
This effect is neglected in the analysis and results in slightly more conservative limits for the doublet model.

For the background samples, several matrix element event generators are combined with parton shower and hadronisation programs. \POWHEG-\textsc{Box} v2~\cite{PowhegBox,Powhegtt,PowhegWt,Powhegst} interfaced to \PYTHIA v6.428~\cite{Pythia6} is used to simulate \ttbar production as well as single-top 
production in association with a $W$ boson, while electroweak $t$-channel single-top quark events are generated using the \POWHEG-\textsc{Box} v1 generator. Events containing $W$ bosons with associated jets (\Wjets) are simulated using the \SHERPA v2.2.0 event generator~\cite{Sherpa,Comix,OpenLoops}, while for diboson events \SHERPA v2.1.1 is used. For the simulation of \ttbar events with additional vector bosons ($\ttbar+V$) the \MGMCatNLO v2.3.2~\cite{Madgraph} event generator interfaced with \PYTHIA~8 is used.

All MC samples are normalised using the highest-order inclusive cross-sections available at $\sqrt{s}=13\,\TeV$.
The dominant backgrounds \ttbar and \Wjets are available at NNLO in QCD~\cite{Czakon:2011xx,Catani:2009sm}. 
The parton distribution function (PDF) sets used are CT10~\cite{CT10} for the \POWHEG-\textsc{Box} and \SHERPA v2.1.1 generators, NNPDF 3.0 NNLO~\cite{NNPDF3} for \SHERPA v2.2.0, and NNPDF 2.3~LO~\cite{NNPDF2} for \PROTOS and \MGMCatNLO. The samples produced with \PROTOS, \MGMCatNLO and \POWHEG-\textsc{Box} use \textsc{EvtGen} v1.2.0~\cite{EvtGen} for the modelling of $b$-hadron decays. 
All simulated samples are processed with the full simulation of the ATLAS detector~\cite{SOFT-2010-01} using \textsc{Geant}~4~\cite{Geant}, and 
with varying numbers of simulated minimum-bias interactions generated with \PYTHIA~8 overlaid on the hard-scattering event to account for the multiple $pp$ interactions in the same or nearby bunch crossings (pile-up).  
The average number of interactions per bunch crossing in simulation matches the distribution in data.
Additional details of the simulation set-up can be found in Refs.~\cite{ATL-PHYS-PUB-2016-002,ATL-PHYS-PUB-2016-003,ATL-PHYS-PUB-2016-004,ATL-PHYS-PUB-2016-005}. 

\section{Event reconstruction and object selection}
\label{event_reco}

Events are required to have at least one vertex candidate with at least two tracks with $\pt > 500\,\mev$.
The primary vertex is taken to be the vertex candidate with the largest sum of squared transverse momenta of all associated tracks.
Two sets of quality and kinematic requirements are imposed on physics objects in this analysis, labelled baseline and signal requirements, 
where the latter label describes a tighter selection than the former.

Electron candidates are reconstructed from cell clusters in the electromagnetic calorimeter which are matched to ID tracks. Baseline electrons are required to have $\pT > 7$\,\GeV, $|\eta| < 2.47$, and satisfy `VeryLoose' likelihood identification criteria, 
defined following the methodology described in Ref.~\cite{ATLAS-CONF-2016-024}.
Signal electrons must pass all baseline requirements, 
have $\pT > 28$\,\GeV, and satisfy the `Tight' likelihood identification criteria~\cite{ATLAS-CONF-2016-024}. 
In addition, the distance of closest approach to the beam in the transverse plane ($d_0$) and the longitudinal distance of this point 
to the primary vertex along the beam direction ($z_0$) 
are required to satisfy $| z_0 \sin \theta| < 0.5$\,mm and $|d_0|/\sigma_{d_0} < 5$, where $\sigma_{d_0}$ is the uncertainty in  $d_0$.
Furthermore, signal electrons must be isolated, requiring that the scalar \pt sum of all tracks in a cone of size $\Delta R < 10\,\GeV / \pt^{e}$ around the electron momentum, with a maximum cone size of 0.2, is lower than 6\% of the electron \pt. 

Muon candidates are reconstructed from combined tracks that are formed from ID and MS tracks, ID tracks matched to MS track segments, standalone MS tracks, or ID tracks matched to a calorimeter energy deposit compatible with a minimum-ionising particle (referred to as calo-tagged muons)~\cite{PERF-2015-10}. Baseline muons are required to have $\pt>6$\,\GeV, $|\eta|<2.7$, and satisfy the `Loose' identification criteria~\cite{PERF-2015-10}.
Signal muons must satisfy the `Medium' identification criteria, are required to have $\pt > 28$\,\GeV, and impact parameters $| z_0 \sin \theta| < 0.5$\,mm and $|d_0|/\sigma_{d_0} < 3$.
In addition, they must be isolated, requiring that the scalar \pt sum of all tracks in a cone of size $\Delta R < 10\,\GeV / \pt^{\mu}$ around the muon momentum, 
with a maximum cone size of 0.3, is lower than 6\% of the muon \pt.

Jet candidates are reconstructed from topological clusters~\cite{PERF-2014-07,PERF-2016-04}, 
built from energy deposits in the calorimeters calibrated to the electromagnetic scale, using the anti-$k_t$ algorithm with radius parameter $R=0.4$~\cite{AntiKt}. 
Baseline (signal) jets are required to have $\pt > 20$\,\GeV ($\pt > 25$\,\GeV) and $|\eta|<4.4$ ($|\eta|<2.5)$.
To reject jets originating from pile-up collisions, signal jets with $\pt < 60\,\GeV$ and $|\eta| < 2.4$ are required to satisfy jet vertex tagger criteria~\cite{PERF-2014-03}.
In order to suppress detector noise and non-collision backgrounds~\cite{DAPR-2012-01}, 0.3\% of the events are vetoed as a baseline jet in an event does not pass the `Loose' jet quality requirements of Ref.~\cite{ATLAS-CONF-2015-029}.
A $b$-tagging algorithm based on multivariate techniques is used to identify jets containing $b$-hadrons, so-called $b$-jets~\cite{PERF-2012-04,ATL-PHYS-PUB-2016-012}. The working point used provides an average tagging efficiency of 77\% for $b$-jets in simulated \ttbar events, as well as a rejection factor of about 130 for light-quark flavour ($u,d,s$) and gluon jets and about 6 for charm jets.

Jets and associated tracks are also used to identify hadronically decaying $\tau$ leptons using the `Loose' identification criteria described in Ref.~\cite{xATL-PHYS-PUB-2015-045}. The $\tau$ candidates are required to have one or three associated tracks, with total electric charge opposite to that of the selected electron or muon, $\pt>20$\,\GeV, and $|\eta|<2.5$. 

The missing transverse momentum is reconstructed from the negative vector sum of the transverse momenta of baseline electrons, muons, jets, and a soft-term built from high-quality tracks that are associated with the primary vertex but not with the baseline physics objects~\cite{ATL-PHYS-PUB-2015-023,ATL-PHYS-PUB-2015-027}.

Energy deposits reconstructed as two different particle candidates 
are resolved by
removing one of the candidates 
according to the following procedure:

\begin{enumerate}
	\item Electron/jet: if a baseline electron and a baseline jet are found within $\DeltaR < 0.2$ the overlapping jet is removed.
	
	\item Muon/jet: if a baseline muon 
overlaps with a baseline jet within $\DeltaR < 0.4$ 
the overlapping jet is removed if
the jet has fewer than three tracks with $\pt > 500\,\mev$.
	\item Jet/lepton:
	\begin{itemize}
		\item If a jet that passes the previous steps overlaps with a muon in a cone of size $\Delta R=0.04+10\,\GeV/\pt^\mu$, up to a maximum size of 0.4, the muon is removed.
		
		\item If a jet that passes the previous steps overlaps with an electron in a cone of size $\Delta R=0.4$, the electron is removed.
	\end{itemize}

	\item Electron/$\tau$: if an electron that passes the previous steps overlaps with a $\tau$ candidate in a cone of size $\Delta R=0.1$, the $\tau$ is removed.

\end{enumerate}

Large-radius jets are 
constructed from signal 
jets using the anti-$k_t$ algorithm with $R=1.0$. Constituent small-radius jets with \pt less than 5\% of the large-radius jet \pt are removed, mainly in order to reduce the impact of soft radiation.


\section{Event selection and background estimation}
\label{sec:signal_region}

Events are required to have 
exactly one signal lepton ($L\!=\!e, \mu$), at least four 
signal jets and 
\mbox{$\met\!>\!300\,\GeV$}. A veto on events with a second lepton, fulfilling the baseline requirements, is used to suppress \ttbar events with two leptons in the final state
(dileptonic \ttbar). Further event selection is performed using objects fulfilling the signal requirements described in the previous section.  
The azimuthal angles between the missing transverse momentum vector $\vec{E}_{\rm T}^{\rm miss}$ and both the leading ($j_{1}$) and sub-leading ($j_{2}$) jets, ordered in \pt, must satisfy the condition $|\Delta\phi(j_i,\vec{E}_{\rm T}^{\rm miss})| > 0.4$ with $i\in\left\{1,2\right\}$, which rejects events with \met arising from mismeasured jets.
In addition, at least one $b$-tagged jet is required.

After applying these requirements, the dominant backgrounds are 
single-lepton plus jets ($1L$) $t\bar{t}$ events, where the lepton originates either directly from the $W$ decay or from a subsequent
$\tau$ decay, 
and \Wjets production. This can be seen in 
Figure~\ref{fig:presel}, which shows a comparison of data and SM expectation as a function of \met and \mt at the so-called preselection level as defined in Table~\ref{tab:cr-def}.
Here \mt is defined as the 
transverse mass\footnote{The transverse mass $\mt$ is defined as $\mt = \sqrt{2p_{\text{T}}^{\text{lep}}\met[1-\cos(\Delta\phi)]}$, where $\Delta\phi$ is the azimuthal angle between the lepton and the missing transverse momentum direction 
and $p_\text{T}^\text{lep}$ is the transverse momentum of the charged lepton.} of the signal lepton and the 
missing transverse momentum.
Requiring \mt to be well above the $W$ boson mass strongly reduces 
$1L$ \ttbar events and \Wjets production as can be seen from the right panel of Figure~\ref{fig:presel}.
The remaining events at large \mt are mostly from \ttbar production where both $W$ bosons decay leptonically.
The two cases of either two leptons ($2L$)
and one lepton and a hadronically decaying $\tau$ ($1L \, 1\tau$) are shown separately.

\begin{figure}[htb]
	\centering
	\includegraphics[width=.5\textwidth]{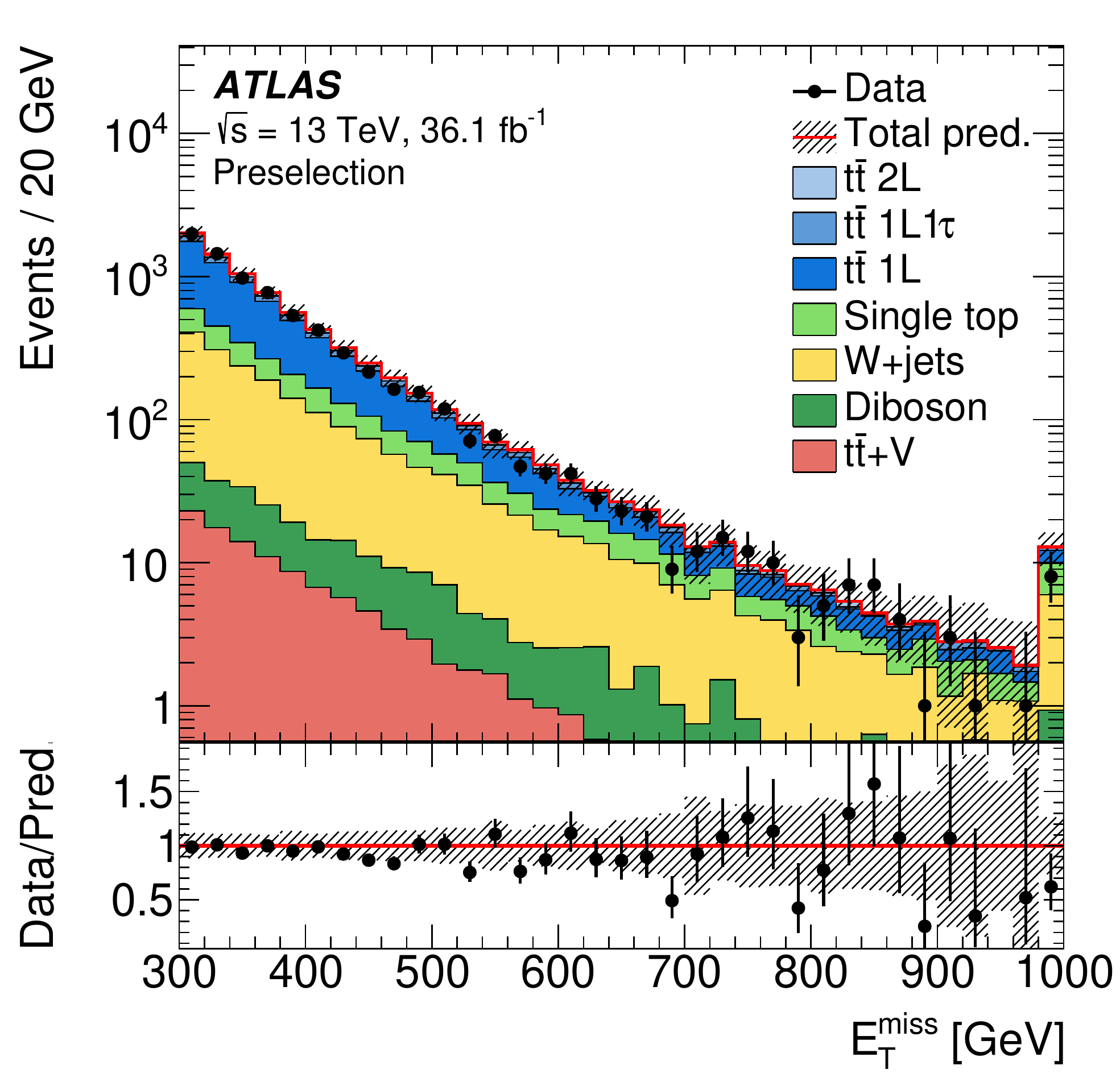}%
	\includegraphics[width=.5\textwidth]{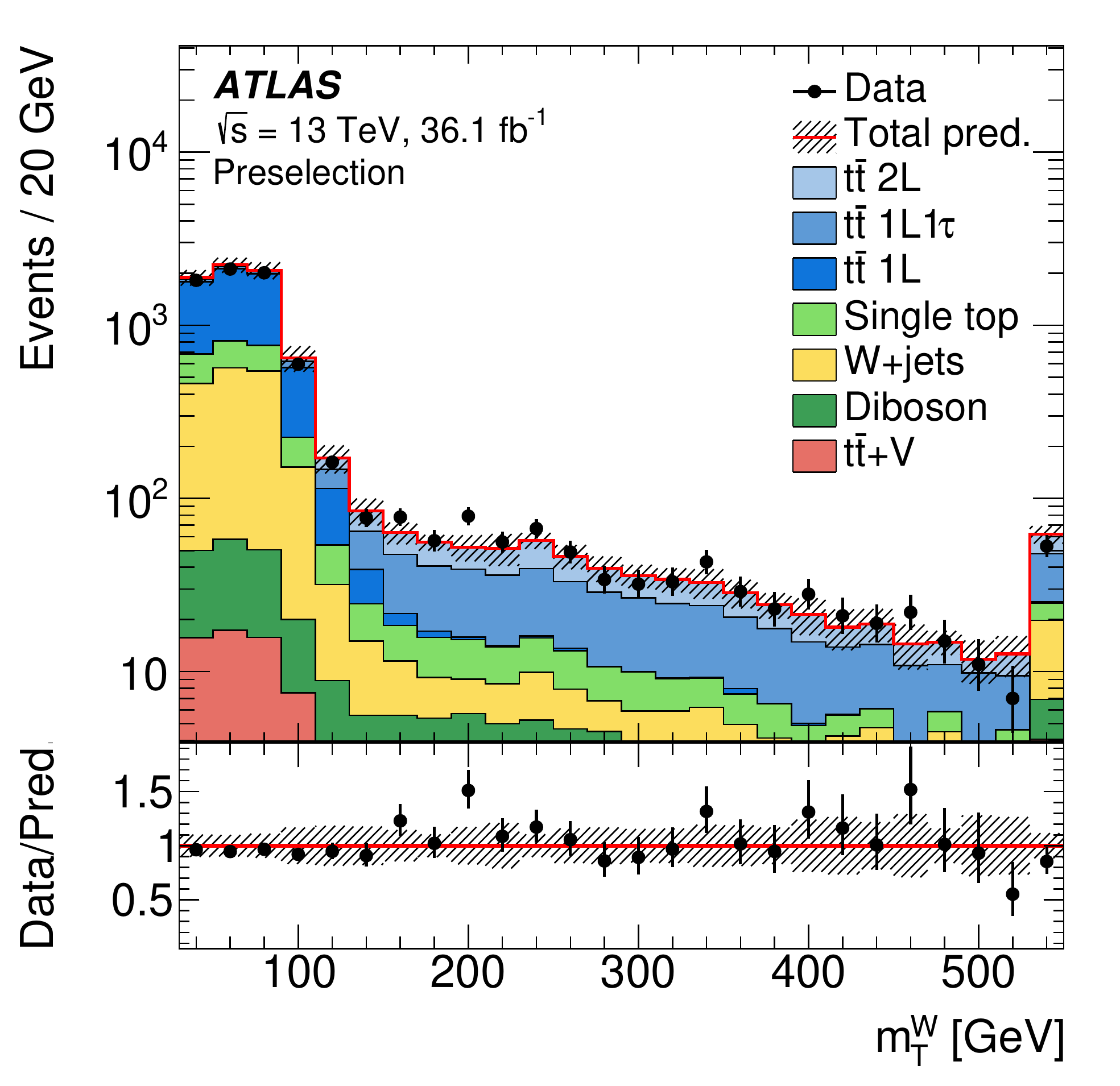}
	\caption{Comparison of data and prediction in the \met distribution (left) and the \mt distribution (right) after the preselection. The lower panels show the ratio of the data to the prediction. The error bands include statistical and systematic uncertainties. The last bin contains the overflow events.}
	\label{fig:presel}
\end{figure}

Additional variables are used to reduce these dileptonic \ttbar backgrounds.
The 
\mtTwo variable~\cite{Lester:1999tx} is a generalisation of the transverse mass applied to signatures where two or more particles are not directly detected~\cite{MT2,AMT2}, defined as
\begin{align*}
	\mtTwo \equiv\underset{\vec{q}_{\text{T}a}+\vec{q}_{\text{T}b}= \vecmet}{\min} \left\{\max (m_{\text{T}a},m_{\text{T}b})\right\}.
\end{align*}
In this formula, $m_{\text{T}a}$ and $m_{\text{T}b}$ are transverse masses calculated using 
two sets of one or more visible particles, denoted
$a$ and $b$, respectively, and 
all possible 
combinations of missing transverse momenta $\vec{q}_{\text{T}a}$ and $\vec{q}_{\text{T}b}$, with $\vec{q}_{\text{T}a}+\vec{q}_{\text{T}b}= \vecmet$.
This analysis uses two specific choices for the sets of visible particles $a$ and $b$, and the corresponding variables are referred to 
as \amtTwo and \mtTwoTau.
The \amtTwo variable is used to suppress dileptonic \ttbar events where one lepton goes undetected.
Thus, the sets of visible particles for the \amtTwo are combinations of the identified lepton and the two jets with the highest $b$-tagging weights, where the lepton is combined with either jet 
and the combination with the lowest resulting \amtTwo is taken.
For dileptonic \ttbar events \amtTwo has a kinematic endpoint at the top quark mass but is expected to extend much higher for processes with additional sources of \met. 
The \mtTwoTau variable targets \ttbar events where one of the $W$ bosons decays via a hadronically decaying $\tau$. It is only constructed in events where a hadronically decaying $\tau$ candidate is found. The $\tau$ candidate and the signal lepton are used as the two visible particles. 
For these events, \mtTwoTau is required to be greater than 80\,\GeV\
as it has a kinematic endpoint at the $W$ boson mass for \ttbar events. 
If fulfilled, 
the $\tau$ candidate is no longer used and instead the jet candidate, from which the $\tau$ is identified, is considered for the rest of the computations. 

\begin{table}[htb]
	\centering
	\caption{Overview of the event selections for the signal region (SR) and the background control regions for \ttbar (TCR) and \Wjets (WCR) processes.
For jet \pt and large-radius jet masses the numbers refer to the objects ordered in \pt.
}
	\begin{tabular}{r|c|c|c|c}
		\toprule
		                 Variable & Preselection &                   SR                   & \hspace{1em}TCR\hspace{1em} & \hspace{1em}WCR\hspace{1em} \\ \midrule
		                     \met & $>300\,\GeV$  &              $>350\,\GeV$              &             \multicolumn{2}{c}{$> 300\,\GeV$}             \\
		                      \mt & $>30\,\GeV$  &             $ > 170\,\GeV$             &        \multicolumn{2}{c}{$\in [30,90]\,\GeV$}        \\
		                  \amtTwo & -- &             $ > 175\,\GeV$             &             \multicolumn{2}{c}{$> 100\,\GeV$}             \\
		                  \mtTwoTau  & -- & $ > 80\,\GeV$ &             \multicolumn{2}{c}{$> 80\,\GeV$} \\
		               \HTmissSig & -- &                 $> 12$                 &                  \multicolumn{2}{c}{--}                   \\
		       Jet \pt & $>25\,\GeV$  &         $>120,80,50,25\,\GeV$          &        \multicolumn{2}{c}{$ >120,80,50,25\,\GeV$}         \\
		        $|\Delta\phi(j_i,\met)|$, $i=1,2$ & $> 0.4$  & $> 0.4$   &      \multicolumn{2}{c}{$>0.4$} \\
		\# $b$-tagged jets &$\ge 1$   &                $\ge 1$                 &           $\ge1$            &            $=0$             \\
		 \# large-radius jets & -- &                $\ge 2$                 &                \multicolumn{2}{c}{$\ge 2$}                \\
		       Large-radius jet mass &-- &             $>80,60\,\GeV$             &            \multicolumn{2}{c}{$>80,60\,\GeV$}             \\
		        \multirow{2}{*}{Large-radius jet \pt} & \multirow{2}{*}{--} & $>290\,\GeV$ if $\met < 450\,\GeV$ &     \multicolumn{2}{c}{ \multirow{2}{*}{$>200\,\GeV$}} \\
		                          & & $>200\,\GeV$ if $\met > 450\,\GeV$ & \multicolumn{2}{c}{} \\ \bottomrule 
	\end{tabular}
	\label{tab:cr-def}
\end{table}

The \HTmissSig~\cite{Nachman:2013bia} variable is used to reduce the number of events with mis-reconstructed \met. 
It is defined as $\HTmissSig \equiv (\HTmiss - 100\,\GeV) / \sigma_{\HTmiss}$, where the exact value of the constant has no 
impact on this analysis and was optimised 
for the 
search mentioned above~\cite{SUSY-2013-15}.
Here \HTmiss is the magnitude of the 
vectorial sum of the signal lepton and jets transverse momenta
and $\sigma_{\HTmiss}$ is the approximate resolution of \HTmiss, computed using the per-event jet energy resolution~\cite{ATL-PHYS-PUB-2015-015}.

\subsection{Signal region selection}

The signal region selection is optimised to target a benchmark signal with a $T$ quark mass of 1\,\TeV, a branching ratio for the $T\rightarrow Zt$ decay of 80\%, and equal branching ratios for the other two decay modes ($T\rightarrow Ht$, $T\rightarrow Wb$). 
This benchmark signal was chosen due to its
mass being at the expected sensitivity reach of this analysis, and its  
high branching ratio into the required $Zt$ decay mode, while at the same time allowing for decays into other final states.  
A summary of the SR selection is given in Table~\ref{tab:cr-def}. 
Due to the invisible decay
of the $Z$ boson, 
a high-\met requirement of at least 350\,\GeV is set. As discussed above, the \mt and \amtTwo variables are used to reduce the
\ttbar and \Wjets backgrounds.
The properties of large-radius jets are used to exploit the structure of the high-mass VLT decay.
At least two large-radius jets with high \pt and large mass are required. These target a hadronically decaying top quark, or the subsequent hadronic $W$ boson decay, as well as a second, hadronically decaying massive boson.  
At very high \met the backgrounds are sufficiently reduced to allow looser requirements on the large-radius jet \pt.
For the benchmark signal, $13.4\pm0.5$ events are expected in the SR, with 95\% of them 
containing an invisibly decaying $Z$ boson. This number decreases to 75\% for the singlet model with a $T$ quark mass of 1\,\TeV, 
where $\BRto{Zt}\simeq 25\%$.
The efficiency of the full event selection for the benchmark signal 
is 1\%, taking all decay modes into account.  
For events with $Z\rightarrow\nu\bar\nu$ and 
one leptonically decaying $W$ boson from a top quark decay,
the 
selection efficiency is about 10\%.

\subsection{Background estimation}

The dominant background in the SR is due to \ttbar production, followed by about equal contributions from \Wjets and single-top production. 
In order to derive the normalisation of the \ttbar and \Wjets processes from data, dedicated control regions are defined and referred to as TCR and WCR, respectively.
The background normalisation and a possible signal contribution are determined simultaneously in a maximum-likelihood fit to the event yields in the signal region and the control regions. 

The CRs are defined by modifying the requirement on \mt to a window around the $W$ boson mass, ensuring orthogonality between the CRs and the SR.
Requirements on \met, \amtTwo, \HTmissSig and the large-$R$ jet \pt are loosened in order to increase the statistical power of the CRs. In the \Wjets control region, a veto on $b$-tagged jets is used to enrich the \Wjets contribution and make the selection orthogonal to the \ttbar control region. Table~\ref{tab:cr-def} details the CR selection in comparison to the SR requirements. 
The normalisation factors, obtained in a fit to the control regions, for \ttbar and \Wjets are $\mu_{\ttbar} = 1.05\pm0.17$ and $\mu_{\Wjets}=0.70\pm0.10$, where the error includes both the statistical and systematic uncertainties.  
Applying these normalisation factors, a comparison of data and simulation in the \met and \HTmissSig distributions is shown 
in both CRs 
in Figure~\ref{fig:cr-plots}. 
The normalisation factors, in particular the one for \Wjets, are tested further as described in Section~\ref{sec:VR} and Section~\ref{systematics}.
\begin{figure}[htb]
	\centering
	\includegraphics[width=.45\textwidth]{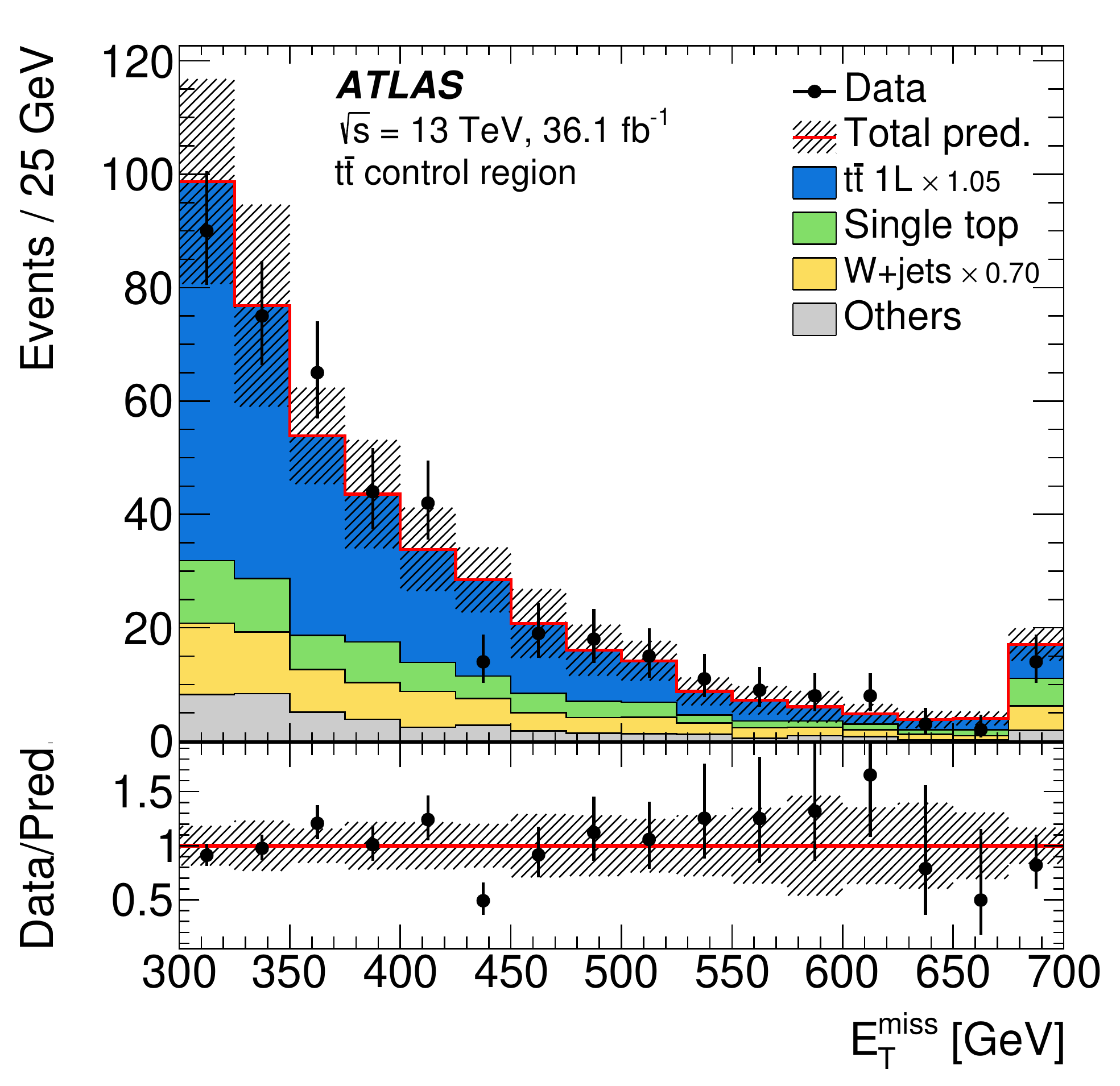}%
	\includegraphics[width=.45\textwidth]{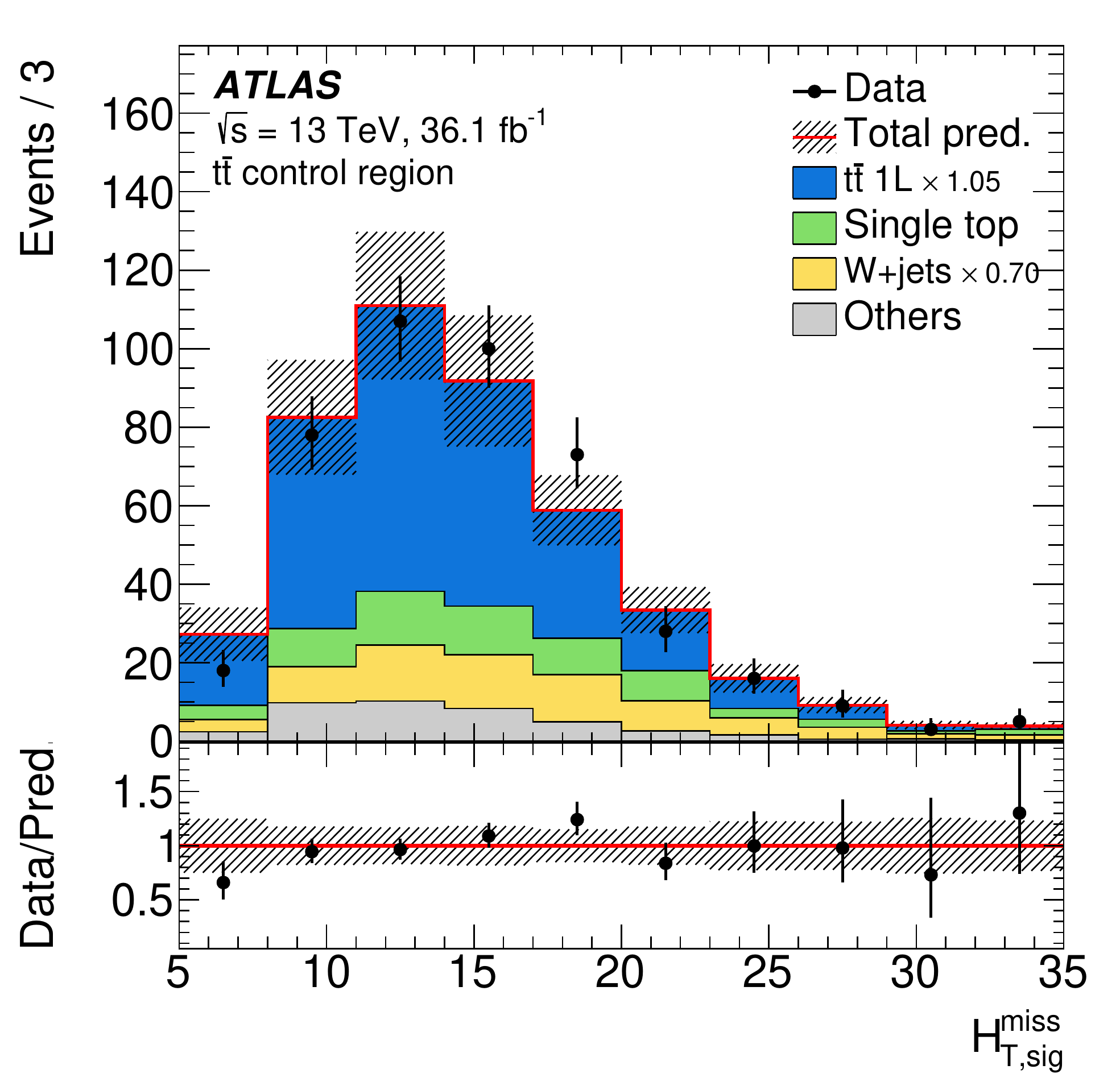}
	\includegraphics[width=.45\textwidth]{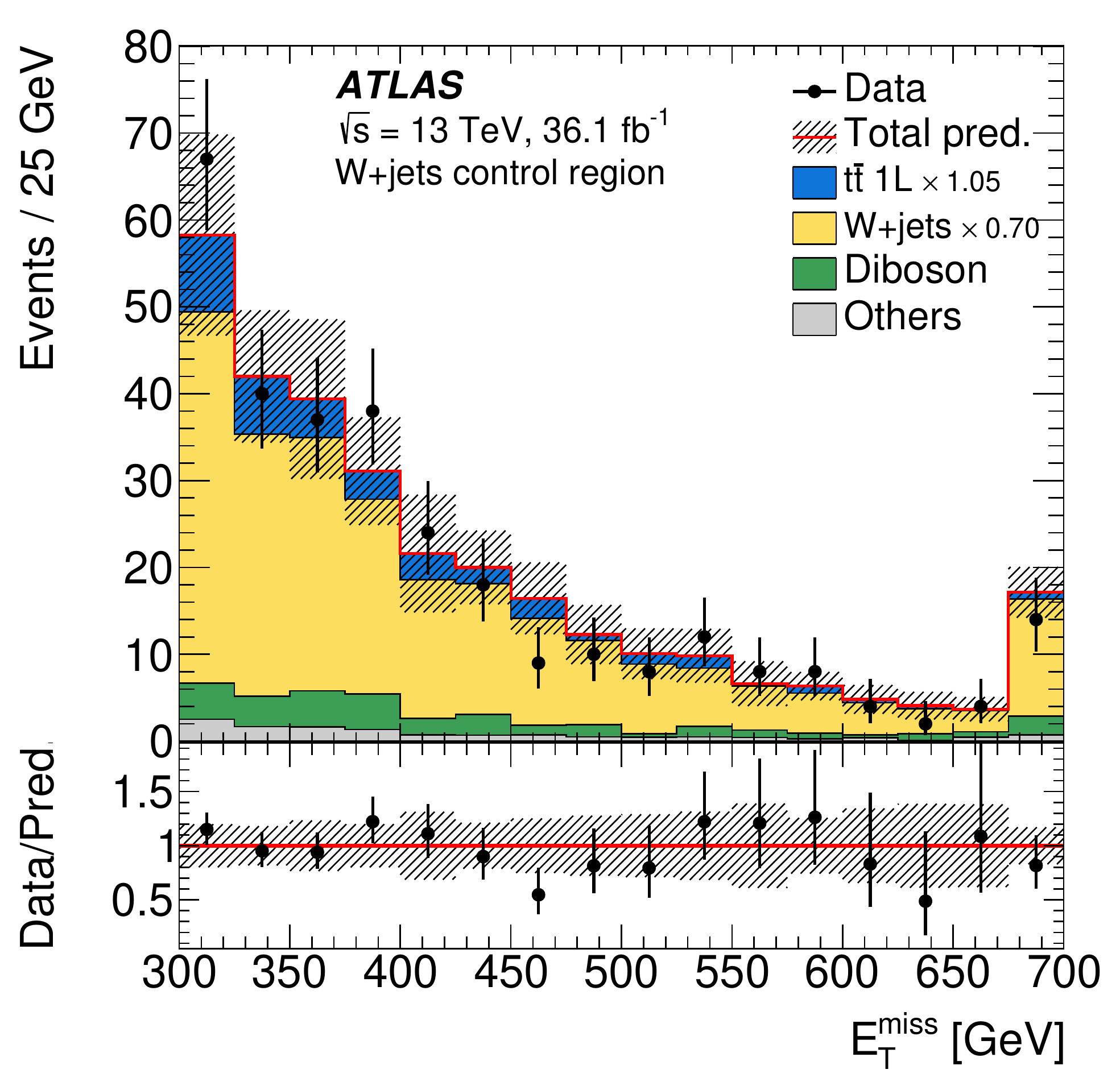}%
	\includegraphics[width=.45\textwidth]{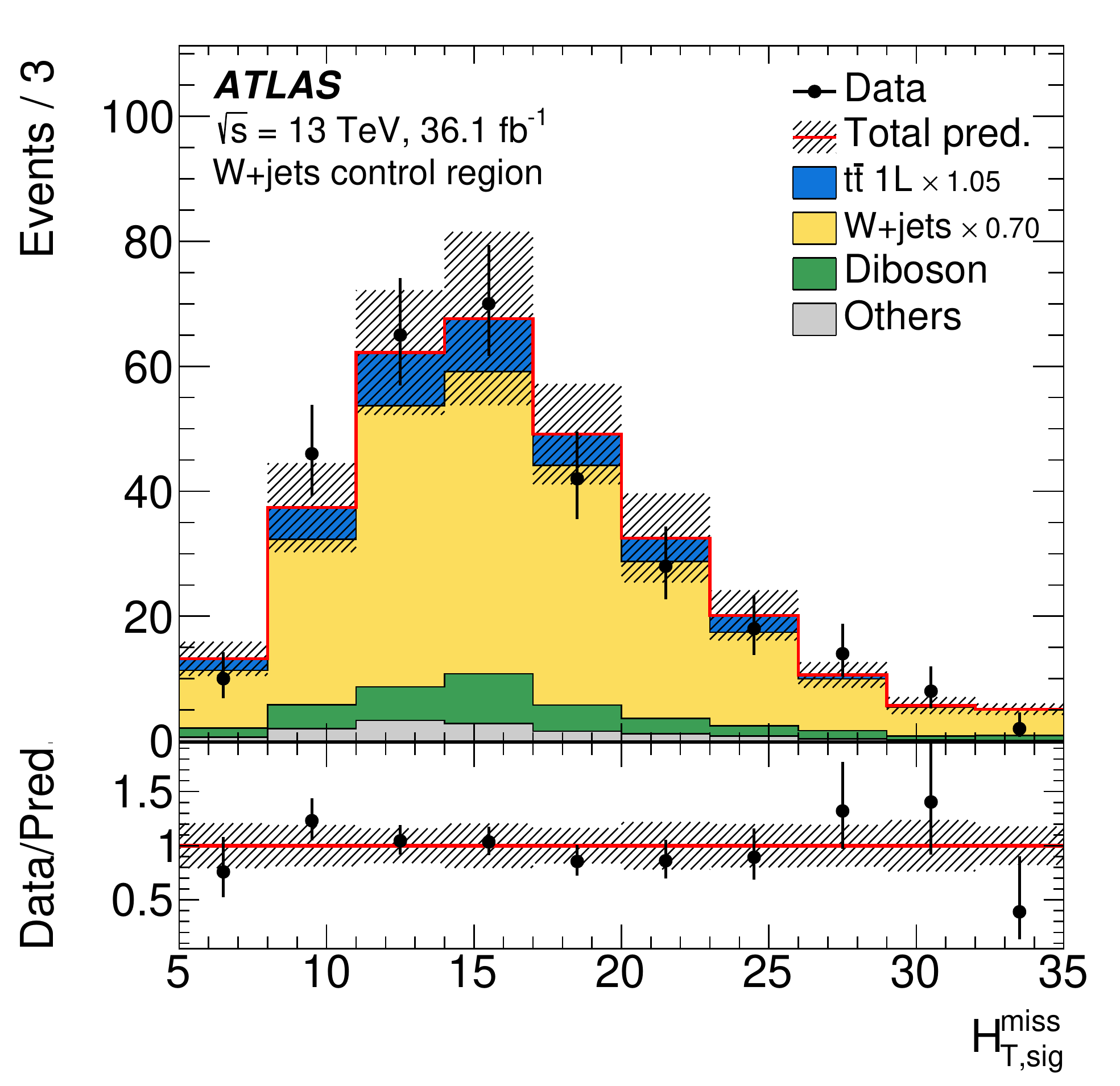}
	\caption{Comparison of data and prediction in the \met distribution (left) and the \HTmissSig distribution (right) in the \ttbar (top) and the \Wjets control regions (bottom). The \ttbar and \Wjets normalisation factors are applied as indicated in the legend. The lower panels show the ratio of the data to the prediction. The error bands include statistical and systematic uncertainties. The last bin contains the overflow events. For the \HTmissSig distributions the first bin also contains underflow events.}
	\label{fig:cr-plots}
\end{figure}

The single-top background is taken from simulation. 
Additional backgrounds considered in the analysis are diboson production as well as $\ttbar$ production in association with a vector boson. These backgrounds make a small contribution to the SR and are also taken from simulation. 
Rare backgrounds, such as the associated production of \ttbar with a Higgs boson, contribute less than 1\% to the total expected background in the SR and are not considered further.
The contribution of multi-jet events is negligible after the requirements on \met and $|\Delta\phi(j_i,\vec{E}_{\rm T}^{\rm miss})|$.

\subsection{Background validation}
\label{sec:VR}
The background estimates are tested in validation regions,
which are not included in the maximum-likelihood fit. The background normalisation factors determined in the CRs are applied in the 
VRs. 
All VRs are defined by requiring exactly one large-radius jet, to ensure orthogonality with the event selection defining the signal and control regions.
Furthermore, some SR requirements are relaxed in order to increase the statistical power.
Three VRs are defined, for the \ttbar, single-top and \Wjets processes. The definitions are summarised in Table~\ref{tab:vr-def}. 
No indication of mismodelling is observed in any of the VRs.
For example, the \met and \mt distributions are shown in Figure~\ref{fig:vr-plots}.

\begin{table}[htb]
	\colorlet{change}{black}
   \centering
	\caption{Overview of the event selections for the \ttbar (TVR), \Wjets (WVR) and single-top (STVR) validation regions, compared to the signal region.
For jet \pt and large-radius jet masses the numbers refer to the objects ordered in \pt.}
	\begin{tabular}{r|c|c|c|c}
		\toprule
		                         Variable &            Signal Region            &     \hspace{1em}TVR\hspace{1em}      &  \hspace{1em}WVR\hspace{1em}  &              STVR              \\ \midrule
		                             \met &            $>350\,\GeV$             &                            \multicolumn{3}{c}{\color{change}$> 300\,\GeV$}                            \\
		                              \mt &           $ > 170\,\GeV$            &           \multicolumn{2}{c|}{\color{change} $>120\,\GeV$}           &  {\color{change} $>60\,\GeV$}  \\
		                          \amtTwo &           $ > 175\,\GeV$            & {\color{change}$\in[100,200]\,\GeV$} & {\color{change} $>100\,\GeV$} & {\color{change} $> 200\,\GeV$} \\
		                        \mtTwoTau &            $ > 80\,\GeV$            &                                   \multicolumn{3}{c}{$> 80\,\GeV$}                                    \\
		                       \HTmissSig &               $> 12$                &                                 \multicolumn{3}{c}{\color{change}--}                                  \\
		                Jet \pt &        $>120,80,50,25\,\GeV$        &                              \multicolumn{3}{c}{$ >120,80,50,25\,\GeV$}                               \\
		$|\Delta\phi(j_i,\met)|$, $i=1,2$ &               $> 0.4$               &                                      \multicolumn{3}{c}{$>0.4$}                                       \\
		                 \# $b$-tagged jets &               $\ge 1$               &                $\ge1$                &    {\color{change}   $=0$}    &    {\color{change} $\ge2$}     \\
		                  \# large-radius jets &               $\ge 2$               &                                \multicolumn{3}{c}{\color{change}$= 1$}                                \\
		               Large-radius jet mass &           $>80,60\,\GeV$            &                                    \multicolumn{3}{c}{$>80\,\GeV$}                                    \\
		                \multirow{2}{*}{ Large-radius jet \pt }& $>290\,\GeV$ if $\met < 450\,\GeV$ &                                   \multicolumn{3}{c}{ \multirow{2}{*}{$>200\,\GeV$} }                                   \\
		                                  & $>200\,\GeV$ if $\met > 450\,\GeV$ &                                         \multicolumn{3}{c}{}                                          \\ \bottomrule
	\end{tabular}
	\label{tab:vr-def}
\end{table}

\begin{figure}[htbp] 
	\centering
	\includegraphics[width=.45\textwidth]{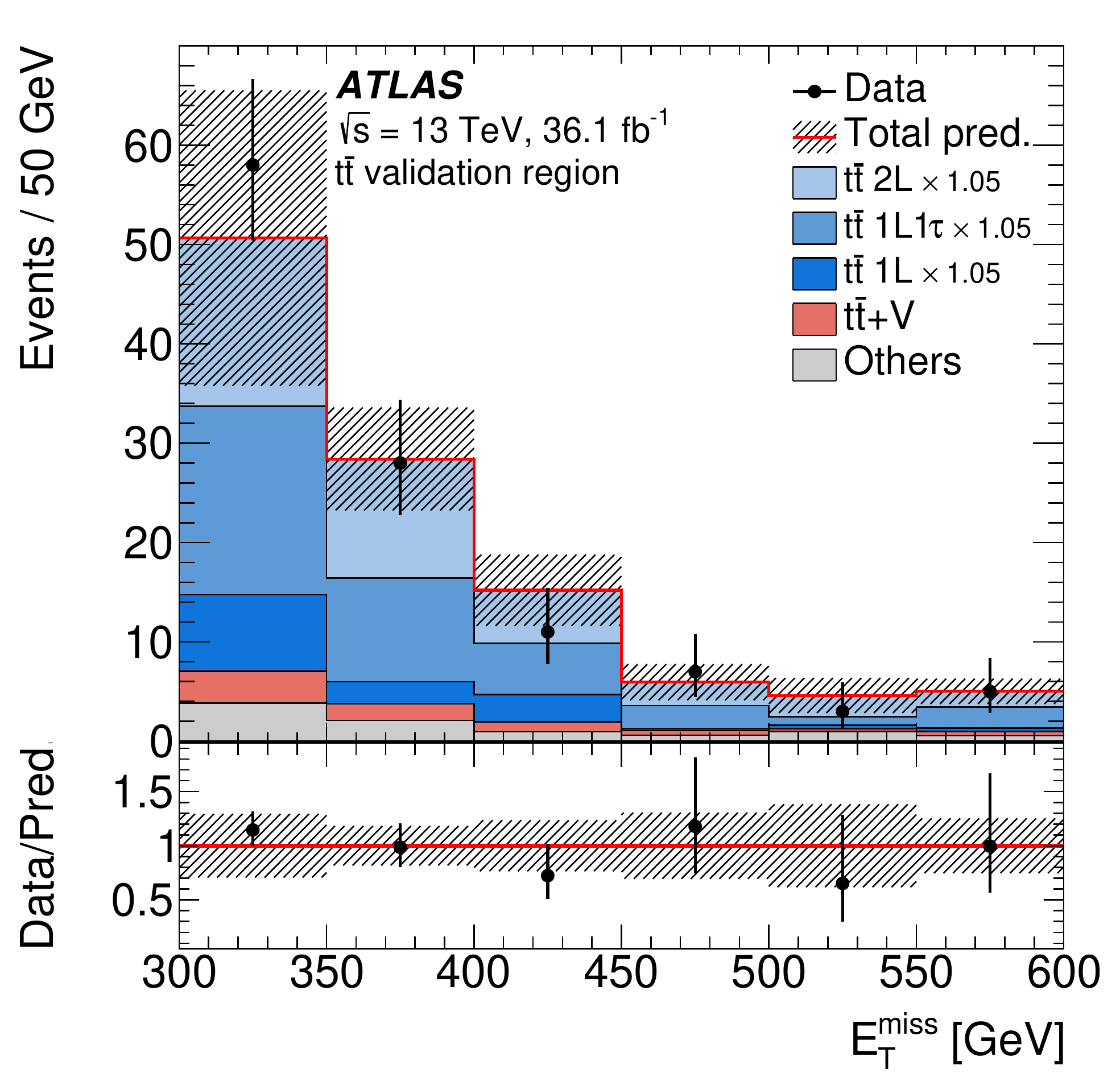}%
	\includegraphics[width=.45\textwidth]{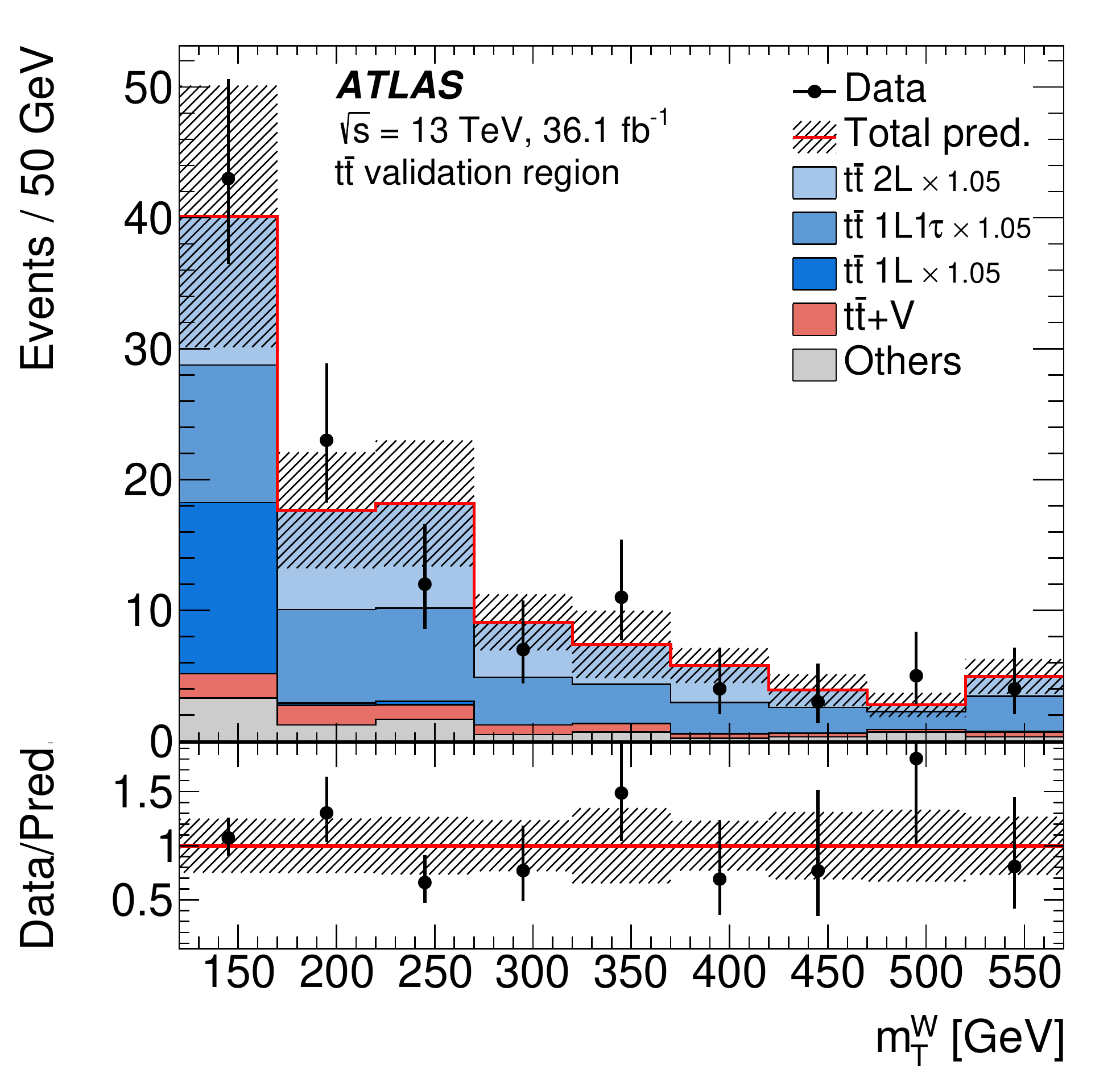}
	\includegraphics[width=.45\textwidth]{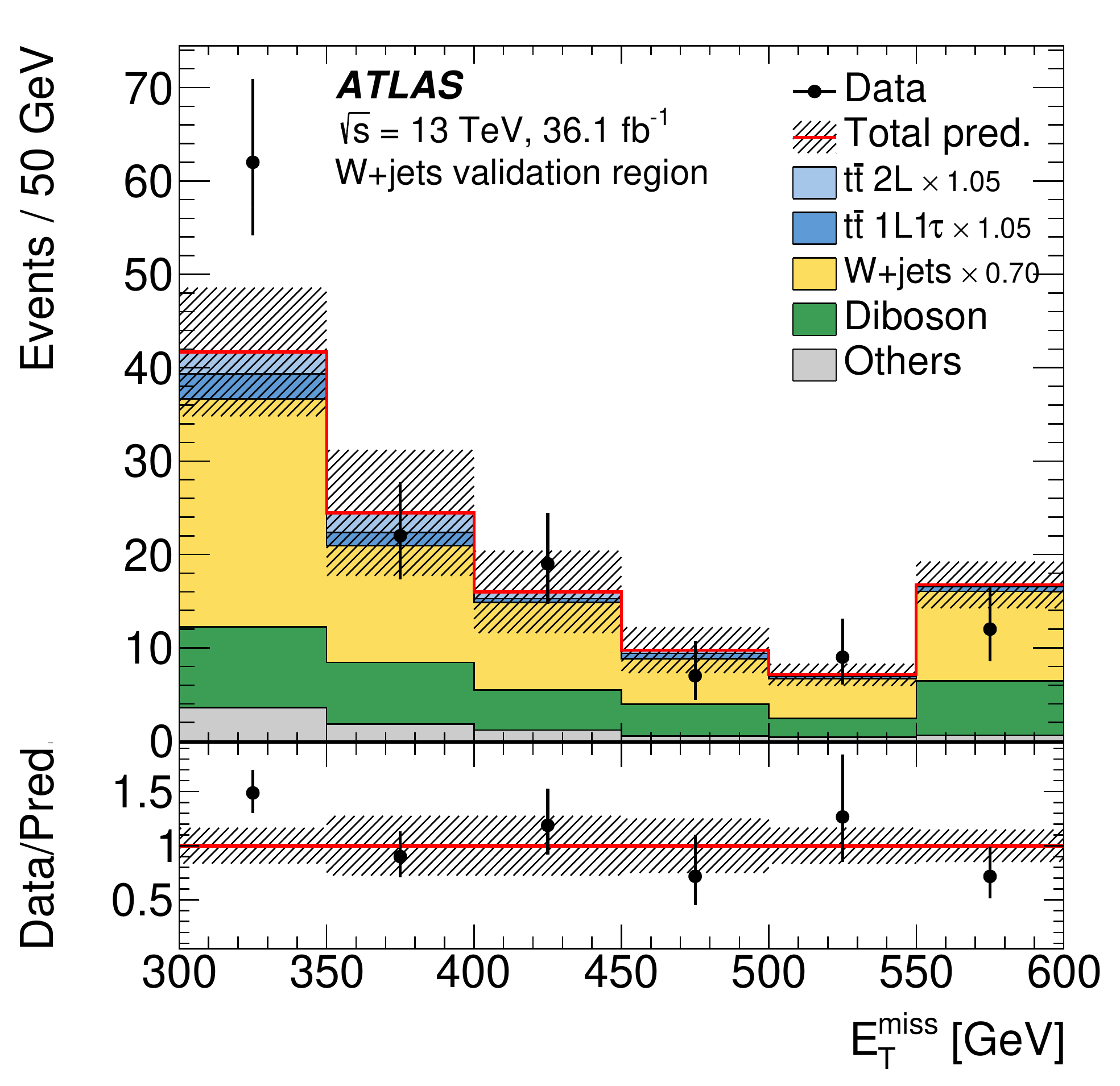}%
	\includegraphics[width=.45\textwidth]{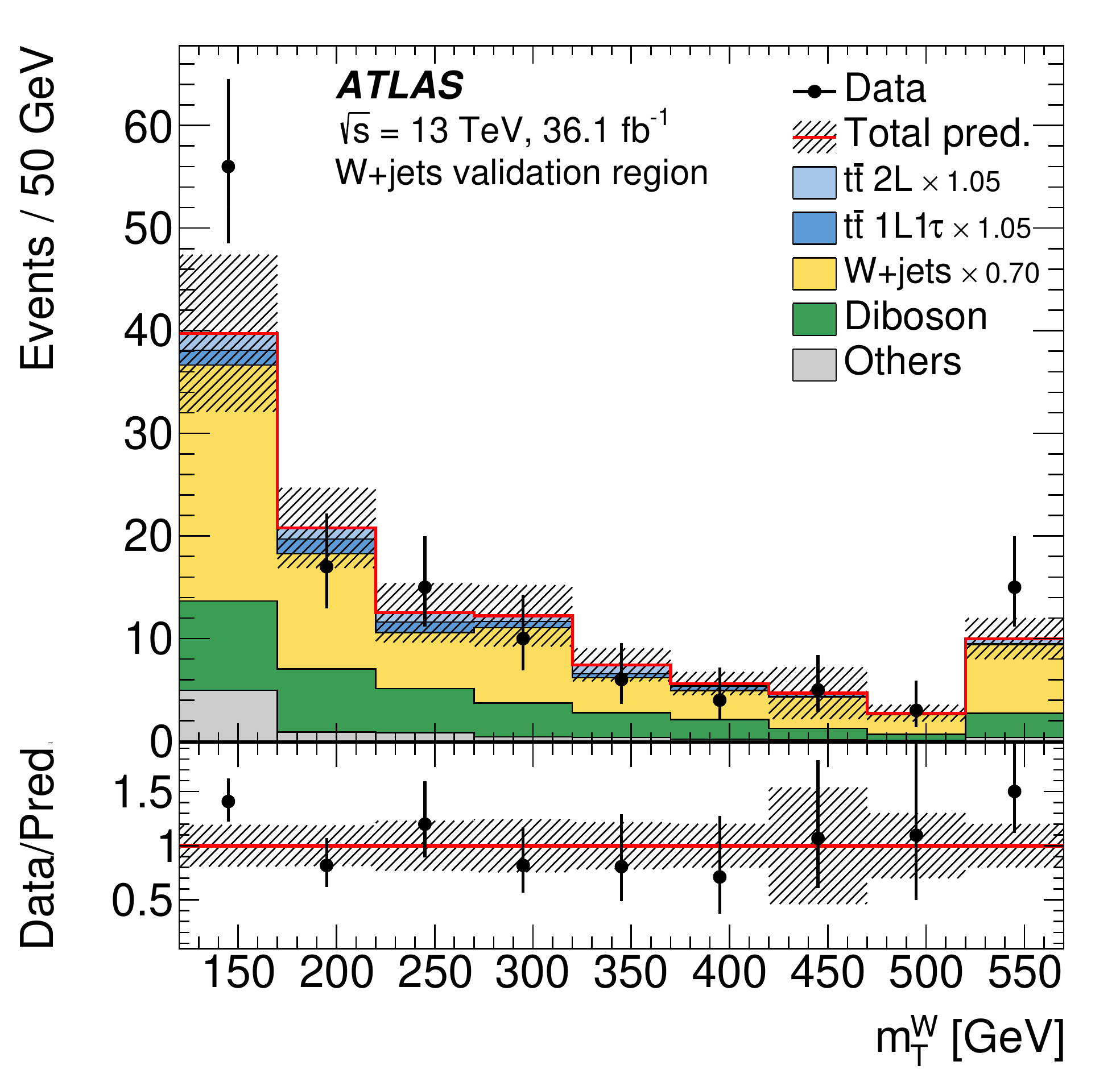}
	\includegraphics[width=.45\textwidth]{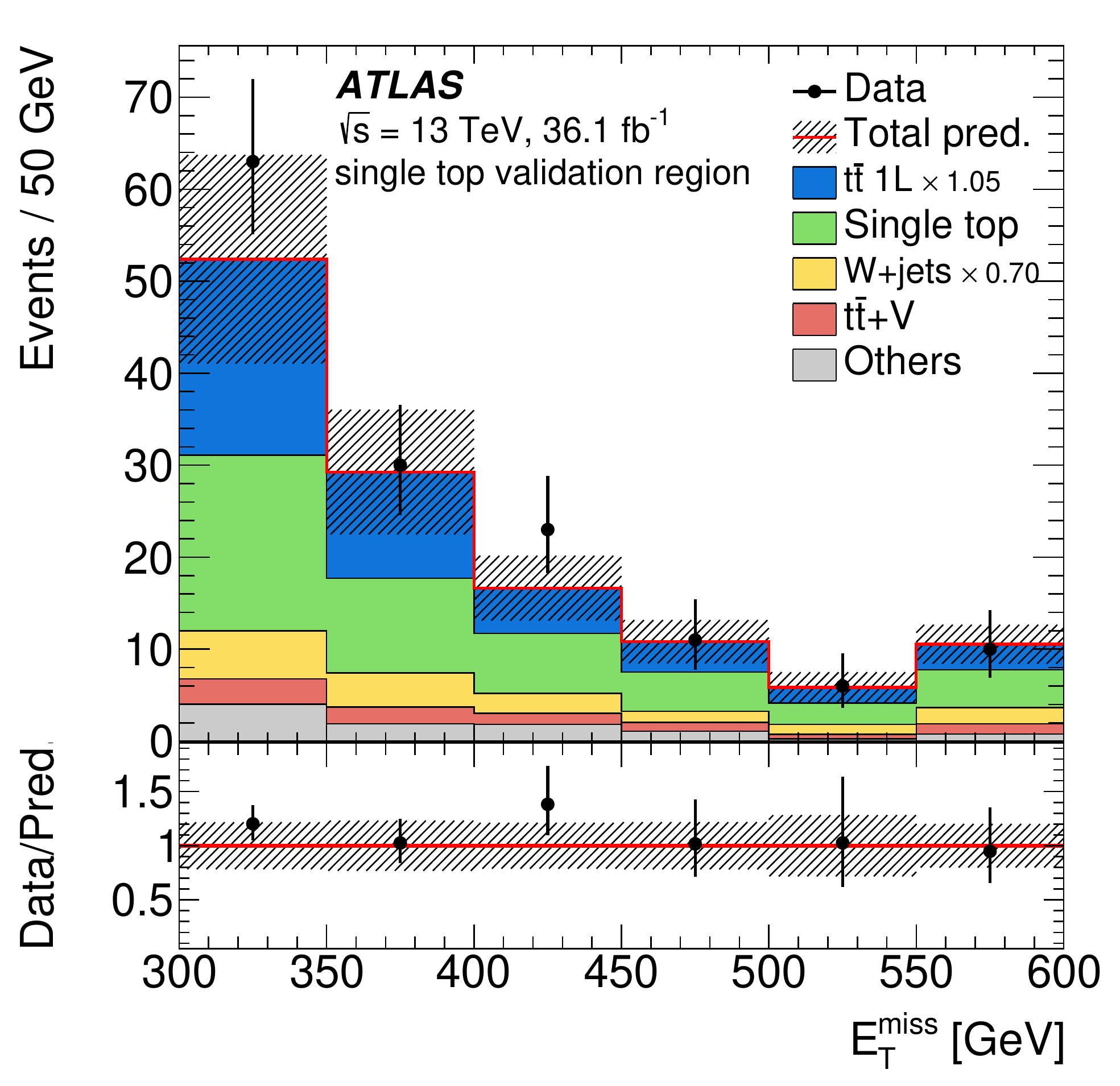}%
	\includegraphics[width=.45\textwidth]{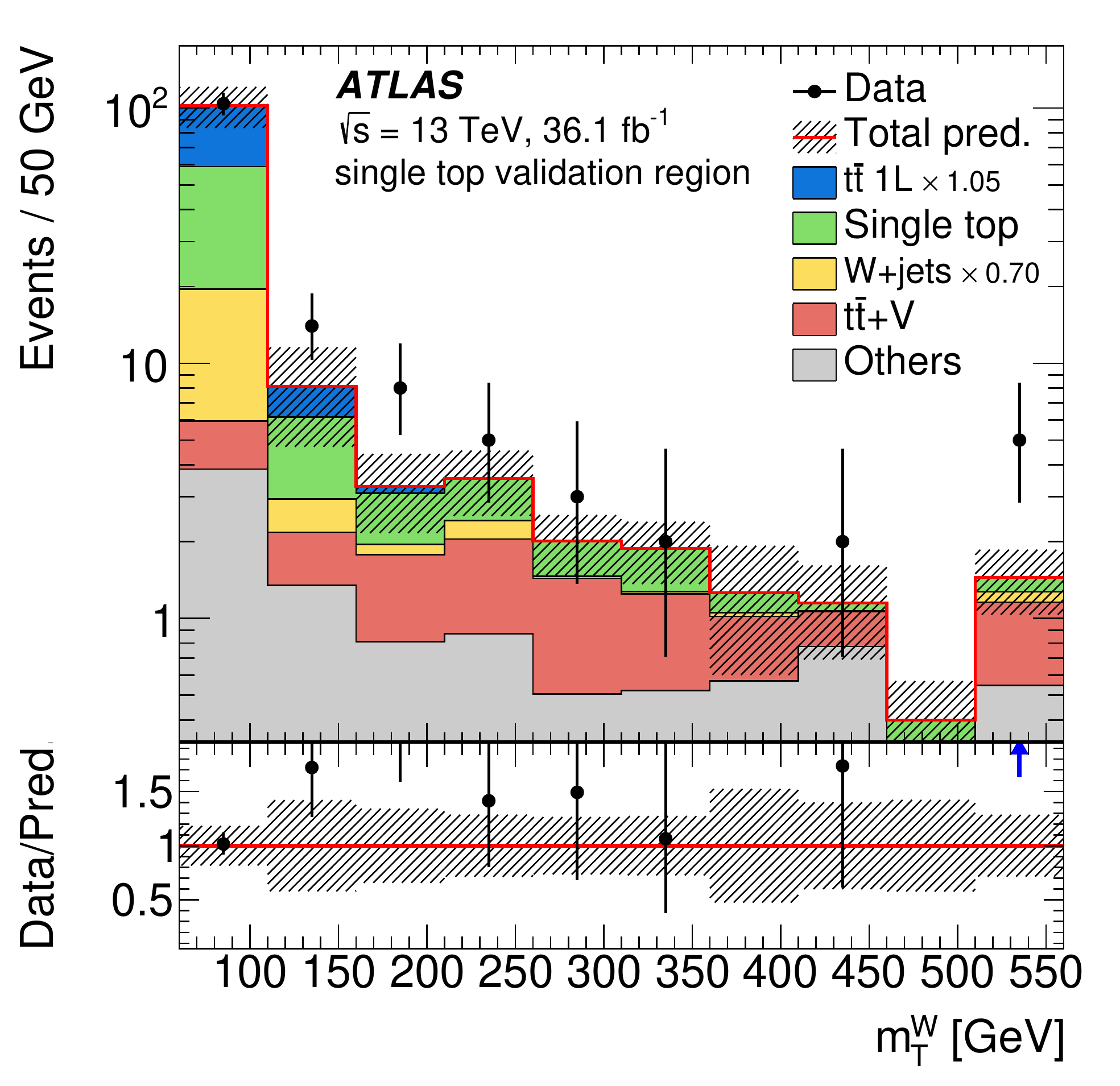}
	\caption{Comparison of data and prediction in the \met distribution (left) and the \mt distribution (right) in the \ttbar (top), \Wjets (middle) and single-top validation regions (bottom). The \ttbar and \Wjets normalisation factors are applied as indicated in the legend. The lower panels show the ratio of the data to the prediction. The error bands include statistical and systematic uncertainties. The last bin contains the overflow events.}
	\label{fig:vr-plots}
\end{figure}



\section{Systematic uncertainties}
\label{systematics}

\begin{table}[htbp]
	\centering
	\caption{Breakdown of the leading systematic uncertainties in the total background prediction in the signal region. The statistical precision is included in uncertainties derived from the comparison of MC samples. The individual uncertainties are correlated, and do not necessarily add in quadrature to the total background uncertainty.}
	\begin{tabular*}{.6\textwidth}{@{\extracolsep{\fill}}lr}
	\noalign{\smallskip}\hline\noalign{\smallskip}
Uncertainty                                           &          SR           \\
	\noalign{\smallskip}\hline\noalign{\smallskip}
Total background prediction                                          &         $6.1$         \\
	\noalign{\smallskip}\hline\noalign{\smallskip}
Total statistical $(\sqrt{n_{\rm exp}})$                             &       $\pm 2.5$       \\
Total background uncertainty                                                                                        & $\pm 1.9\ [31\%] $ \\
	\noalign{\smallskip}\hline\noalign{\smallskip}
\noalign{\smallskip}\hline\noalign{\smallskip}
$\ttbar$ MC generator & $\pm 1.1\ [17\%] $  \\
$\ttbar$ fragmentation                                                                                              & $\pm 0.8\ [14\%] $  \\
$\ttbar$ radiation                                                                                                  & $\pm 0.7\ [11\%] $  \\
MC stat. (nominal samples)                                                                                                           & $\pm 0.7\ [11\%] $  \\
$\ttbar$ Single-top interference                                                                                    &  $\pm 0.6\ [11\%] $  \\
Single-top radiation                                                                                                &  $\pm 0.4\ [6.6\%] $  \\
$\mu_{\ttbar}$                                                                                                      &  $\pm 0.4\ [6.6\%] $  \\
Diboson fact. scale                                                                                                 &  $\pm 0.4\ [6.5\%] $  \\
Diboson renorm. scale                                                                                                &  $\pm 0.4\ [6.1\%] $  \\
\Wjets heavy flavour fraction                                                                                       &  $\pm 0.3\ [5.3\%] $  \\
Jet mass resolution                                                                                                 &  $\pm 0.3\ [5.0\%] $  \\
Diboson resum. scale                                                                                                &  $\pm 0.3\ [4.7\%] $  \\
Flavour-tagging light-jet mistag rate                                                                               &  $\pm 0.3\ [4.5\%] $  \\
Single-top fragmentation                                                                                            &  $\pm 0.2\ [3.5\%] $  \\
Flavour-tagging $c$-jet mistag rate                                                                                 &  $\pm 0.2\ [3.4\%] $  \\
\Wjets MC generator                                                                                                 &  $\pm 0.2\ [3.1\%] $  \\
\ttbar PDF                                                                                                          &  $\pm 0.2\ [2.8\%] $  \\
$\mu_{\Wjets}$                                                                                                      &  $\pm 0.2\ [2.4\%] $  \\
\Wjets merging scale                                                                                                &  $\pm 0.1\ [2.4\%] $  \\
\Wjets renorm. scale                                                                                                &  $\pm 0.1\ [2.3\%] $  \\
\Wjets resum. scale                                                                                                 &  $\pm 0.1\ [2.3\%] $  \\
\Wjets fact. scale                                                                                                  &  $\pm 0.1\ [2.3\%] $  \\
Jet energy scale ($1^{\mathrm{st}}$ component)                                                                                     &  $\pm 0.1\ [2.0\%] $  \\
$\ttbar+V$ cross-section                                                                                            &  $\pm 0.1\ [1.9\%] $  \\
$\ttbar+V$ scale                                                                                                    &  $\pm 0.1\ [1.5\%] $  \\
Jet energy scale ($2^{\mathrm{nd}}$ component)                                                                                     &  $\pm 0.1\ [1.4\%] $  \\
Jet energy resolution                                                                                               &  $\pm 0.1\ [1.3\%] $  \\
Trigger efficiency                                                                                                  & $\pm 0.1\ [1.0\%] $  \\
	\noalign{\smallskip}\hline\noalign{\smallskip}                                                                         &
\end{tabular*}

	\label{tab:syst}
\end{table}

Systematic uncertainties from experimental sources and from the theoretical predictions affect the signal and background estimates.
They are  
included as nuisance parameters with Gaussian constraints and are profiled in the likelihood fits that determine the background normalisation factors and the possible signal contribution. The uncertainties are not constrained further in the fits as the analysis 
has as many bins as free parameters.
The uncertainties in the \ttbar and \Wjets backgrounds in the SR result from the 
relative difference of each systematic source in the SR to the respective CR
due to the normalisation in the fit.
In general the systematic uncertainties are smaller than the statistical uncertainties.

Dominant experimental uncertainties come from imperfect knowledge of the scale and resolution of both jet energies and masses
as well as the modelling of the $b$-tagging efficiency for $b$-jets, $c$-jets and light-flavour jets.
The jet energy scale uncertainty is split into three components~\cite{ATL-PHYS-PUB-2015-014}.
The relative uncertainty in the estimated background yields in the SR from these sources is 1\%--5\%. 
In addition, the uncertainty in the mass scale and the resolution of the small-radius jets is considered. 
It has a small effect on the background, but affects the signal because 
hadronically decaying $Z$ and $W$ bosons can be reconstructed as a single small-radius jet. 
Additional small experimental uncertainties are due to 
the scale and resolution of the \met soft-term, which captures all tracks not associated with calibrated objects in the event, 
the modelling of lepton-related quantities, the trigger efficiency, as well as the uncertainty in the integrated luminosity and the pile-up profile.

The uncertainties in the modelling of the \ttbar 
background come from the choice of MC event generator, 
the hadronisation and fragmentation modelling, and the amount of initial- and final-state radiation. 
The MC generator uncertainty is estimated by comparing events produced with \POWHEG-\textsc{Box} 
and with {\MGMCatNLO},
both in conjunction with \HERWIGpp~\cite{Herwigpp}, and is the dominant uncertainty in the \ttbar modelling, 
amounting to an uncertainty of 17\% in the total background yield in the SR. 
The hadronisation and fragmentation uncertainty is about 13\%, assessed by comparing events generated with \POWHEG-\textsc{Box} and hadronised with either \PYTHIA or \HERWIGpp.
\POWHEG-\textsc{Box}--\PYTHIA samples with different parton-shower radiation
settings, different settings of QCD radiation at NLO accuracy,
and factorisation and renormalisation scales varied by factors of 0.5 and 2~\cite{ATL-PHYS-PUB-2016-004} yield an uncertainty of 11\%.

The uncertainty in the modelling of the \Wjets process is estimated by comparing the nominal simulation to a \MGMCatNLO simulation. 
In addition, the effects of independently varying the scales for the renormalisation, factorisation, and resummation by factors of 0.5 and 2
are taken into account. 
Also, the scale of the matching scheme related to the merging of matrix elements and parton showers is varied between 15\,\GeV and 30\,\GeV. 
Since the \Wjets background is normalised in a CR with a veto on $b$-tagged jets, additional uncertainties of about 30\% in the flavour composition of the \Wjets events are applied in all regions requiring at least one $b$-tagged jet. These uncertainties are based on the measurement reported in Ref.~\cite{STDM-2012-11}, extrapolated to higher jet multiplicities. As a further check, a region with at least one $b$-tagged jet is defined, which is enriched in, though not dominated by, the \Wjets contribution.
Good agreement between data and SM expectation is found when applying the normalisation factors obtained in the CRs.
The total uncertainty in the \Wjets modelling in the SR is about 40\%, resulting in a 6\% uncertainty in the total background estimate.

\FloatBarrier

The dominant uncertainty in the single-top processes stems from the modelling of the interference between the \ttbar and $Wt$ processes at NLO, 
leading to an uncertainty of 11\% in the total background yield in the SR. 
The uncertainty is estimated using inclusive $WW bb$ events, generated using \MADGRAPH at LO, which are compared with the sum of the \ttbar and $Wt$ processes. 
To account for the different orders in \alphas, the number of events in the inclusive $WW bb$ sample is scaled to the sum of the \ttbar and $Wt$ events 
after an event selection requiring only \MET > 200\,\GeV, four jets and exactly one lepton.
Uncertainties from the MC event generator, 
the hadronisation and fragmentation modelling, and the amount of initial- and final-state radiation are estimated in the same 
way as for \ttbar.
Additionally, an uncertainty of 5.3\%~\cite{SingleTopWTXS} in the single-top cross-section is taken into account,
amounting to an uncertainty of less than 1\% in the total background yield in the SR.

Since the diboson and $\ttbar + V$ backgrounds are not normalised in control regions, the analysis is sensitive to their respective cross-section uncertainties. Additionally, the effects of varying the renormalisation, factorisation and resummation scales are taken into account. These uncertainties have a negligible impact on the final results.

The effect of PDF uncertainties on the acceptance for 
the signal and the \ttbar background is included but found to be negligible. A breakdown of the leading systematic uncertainties is summarised in Table~\ref{tab:syst}.


\FloatBarrier

\section{Results}
\label{sec:result}

The number of events observed in the SR, CRs and the VRs is shown in Table~\ref{tab:bkg-only-yields}, together with the backgrounds estimated in the simultaneous fit.
The results are also summarised in the left panel of Figure~\ref{fig:pulls}. Good agreement of data and simulation is observed in the VRs. 
No significant excess above the SM background is found in the SR. 
The \met distribution in the SR for data and SM expectation is presented in the right panel of Figure~\ref{fig:pulls}, 
where the additional contribution
from a $T$ quark signal with a mass of $1.1\,\TeV$ is shown. 
For this particular choice of signal, an additional nine events
would be expected.
The largest signal contribution is expected at about half the  $T$ mass.

\begin{table}[bth]
	\centering
	\caption{Number of events observed in the signal, control and validation regions, together with the estimated SM backgrounds. The normalisation factors determined in the simultaneous fit are applied. The uncertainties include all statistical and systematic sources. The individual uncertainties are correlated, and do not necessarily add in quadrature to the total background uncertainty.}
	\label{tab:bkg-only-yields}
	\setlength{\tabcolsep}{0.0pc}
	{\small
	\begin{tabular*}{\textwidth}{@{\extracolsep{\fill}}lrrrrrr}
\noalign{\smallskip}\hline\noalign{\smallskip}
{\bf Region}           & SR            & TCR            & WCR            & TVR            & WVR            & STVR              \\[-0.05cm]
\noalign{\smallskip}\hline\noalign{\smallskip}
Observed events          & $7$              & $437$              & $303$              & $112$              & $131$              & $143$                    \\
\noalign{\smallskip}\hline\noalign{\smallskip}
Fitted bkg events         & $6.1 \pm 1.9$ & $437 \pm 21$ & $303 \pm 17$ & $109 \pm 35$ & $127 \pm 31$ & $125 \pm 27$     \\
\noalign{\smallskip}\hline\noalign{\smallskip}
        Fitted $t \bar t$ events         & $2.5 \pm 1.7$ & $280 \pm 40$ & $38 \pm 15$ & $90 \pm 40$ & $15 \pm 8$ & $53 \pm 23$     \\
        Fitted $W$ + jets events         & $1.1 \pm 0.7$ & $70 \pm 28$ & $224 \pm 27$ & $3.5 \pm 2.0$ & $77 \pm 30$ & $15 \pm 7$     \\
        Fitted singletop events         & $1.1 \pm 0.7$ & $63 \pm 24$ & $10 \pm 5$ & $4.2 \pm 2.6$ & $3.3_{-3.3}^{+3.5}$ & $46 \pm 17$     \\
        Fitted $t \bar t + V$ events         & $0.91 \pm 0.20$ & $9.7 \pm 1.6$ & $1.03 \pm 0.30$ & $7.0 \pm 1.4$ & $1.9 \pm 0.7$ & $8.3 \pm 1.4$     \\
        Fitted diboson events         & $0.6 \pm 0.6$ & $11 \pm 5$ & $30 \pm 12$ & $1.3 \pm 1.3$ & $31 \pm 9$ & $1.7 \pm 1.1$     \\
 \noalign{\smallskip}\hline\noalign{\smallskip}
MC exp. bkg events              & 6.5 & 450 & 398 & 106 & 160 & 129     \\
\noalign{\smallskip}\hline\noalign{\smallskip}
\end{tabular*}

	}
\end{table}

\begin{figure}[bth]
	\centering
	\includegraphics[width=.5\textwidth]{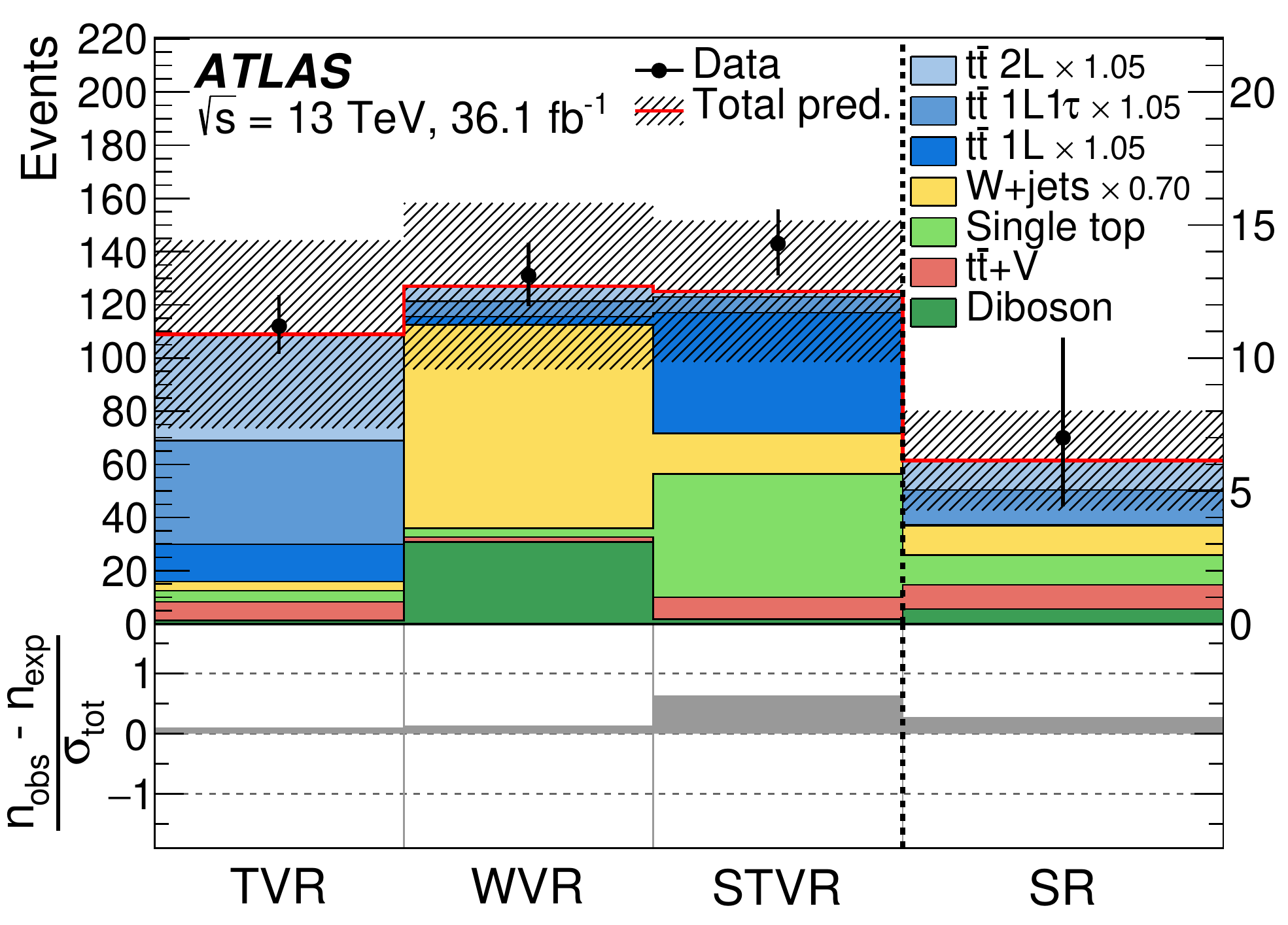}%
	\includegraphics[width=.5\textwidth]{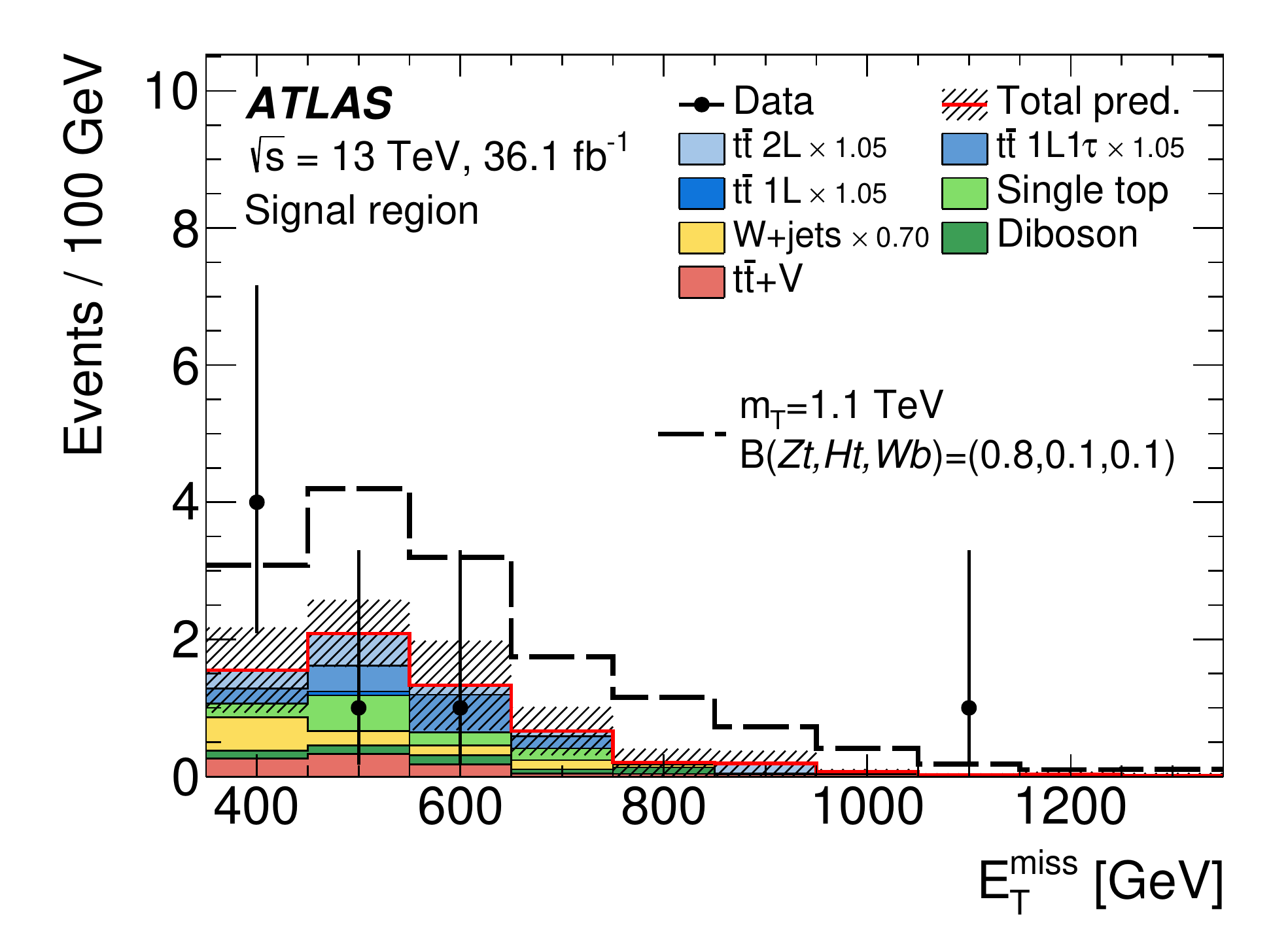}
	\caption{
Left panel: Comparison of the observed data ($n_{\rm obs}$) with the predicted background ($n_{\rm exp}$) in the VRs and SR. 
The bottom panel 
shows the significance of the difference between data and predicted background, where the significance is based on the total uncertainty ($\sigma_{\rm tot}$). 
Right panel: 
Comparison of data and prediction in the \met distribution in the SR.
The error bands include statistical and systematic uncertainties.
	The expected shape for a signal with $m_T=1.1\,\TeV$ and \BRto{Zt} = 80\% 
is added on top of the SM prediction.
}
	\label{fig:pulls}
\end{figure}

Upper limits on the signal cross-section as a function of the VLT mass are derived at 95\% CL using profile-likelihood ratio tests 
following the $CL_s$ prescription~\cite{CLs,Cowan:2010js,HistFitter}. 
They are calculated for the weak-isospin singlet and doublet models, 
which feature mass-dependent branching ratios~\cite{1306-0572}, as well as for pure $T\rightarrow Zt$ decays, using
MC samples generated for 15 different VLT masses between 500\,\GeV and 1.4\,\TeV.
The contribution from additional vector-like quarks, in particular from the $X$ or $B$ quark in the $(X^{5/3}, T)$ or $(T, B)$ doublets, is neglected. 
This results in conservative limits in the case of the doublet model, where the $X$ or $B$ quarks are expected to always decay 
to a $Wt$ final state~\cite{Protos}, 
leading to about 50\% more events in the SR for the case of a $T$ quark mass of 1 TeV and a second $X$ or $B$ quark with the same mass.
The additional contribution in the TCR is less than 10\%.

The cross-section times branching ratio limits as a function of mass are shown in Figure~\ref{fig:xs-limit} 
for the pure $T\rightarrow Zt$ decay, as well as for the singlet and doublet cases.
Comparing the cross-section limits to the theoretical expectation, 
lower limits on the $T$ quark mass can be derived. The limits are summarised in Table~\ref{tab:mass-limit}. 
The sensitivity to the singlet model is lower due to the low 
\BRto{Zt}, which approaches 25\% in the high-mass limit.

\begin{figure}[htbp]
	\centering
	\includegraphics[width=.5\textwidth]{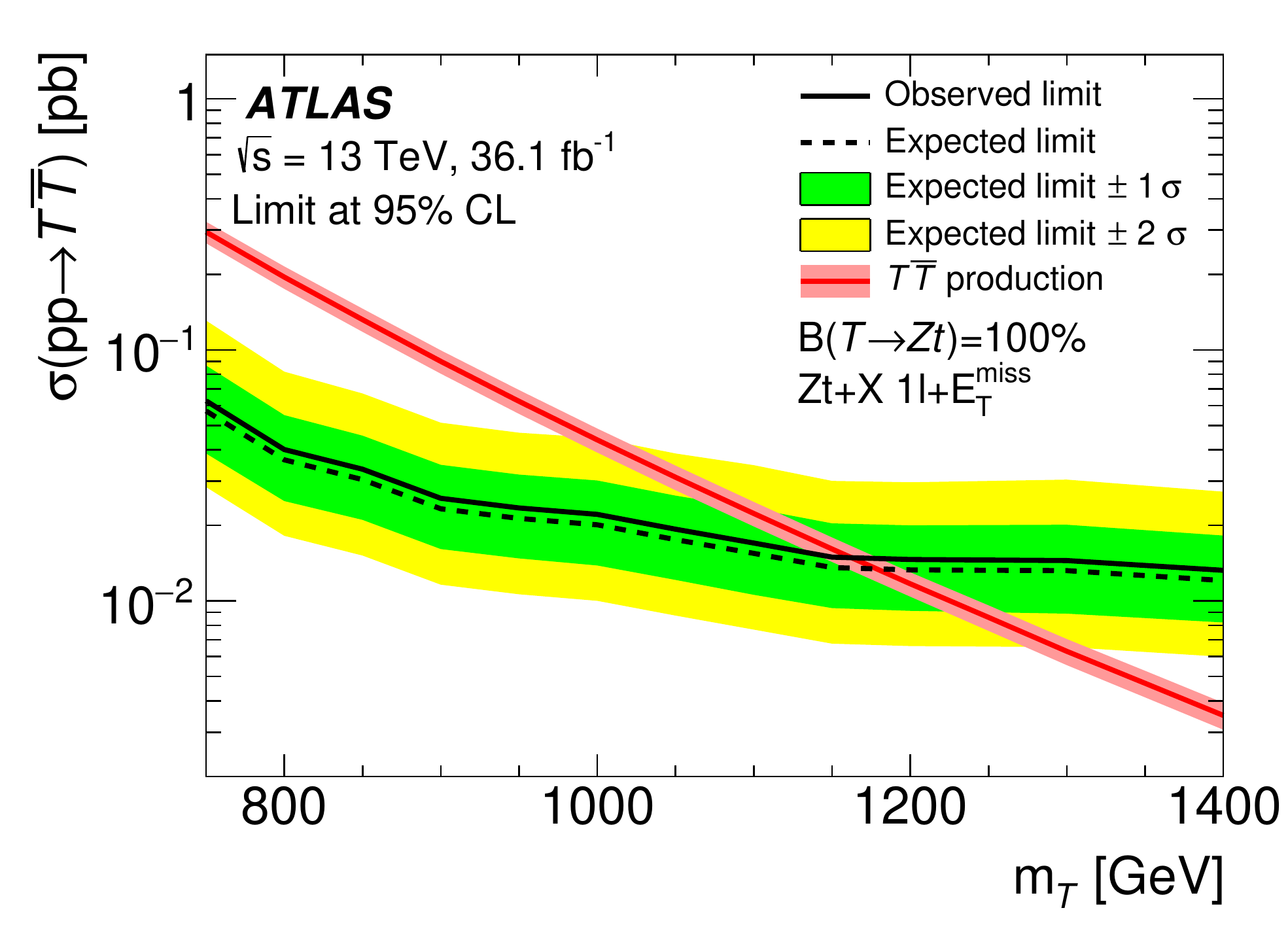}\\
	\includegraphics[width=.5\textwidth]{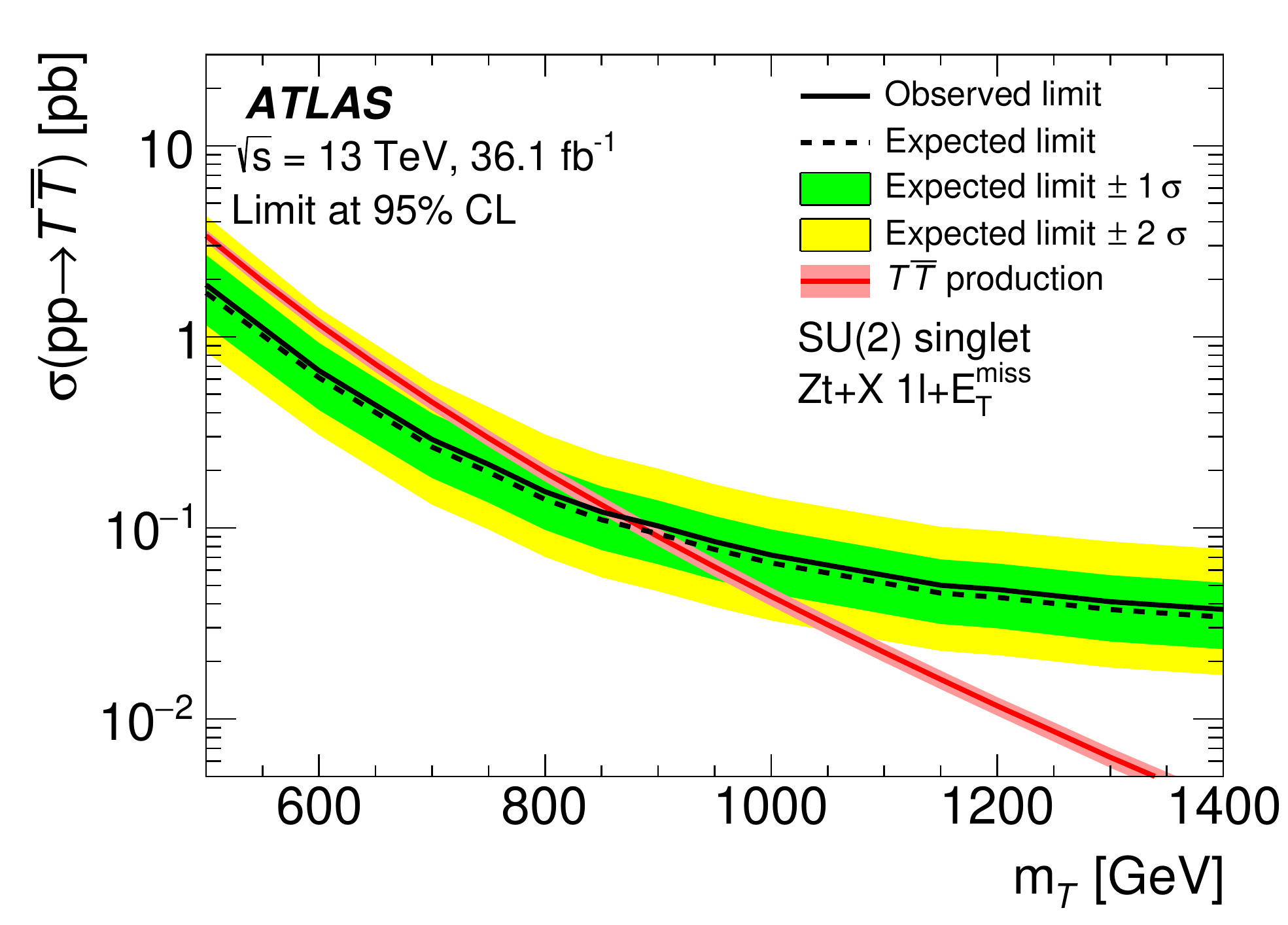}%
	\includegraphics[width=.5\textwidth]{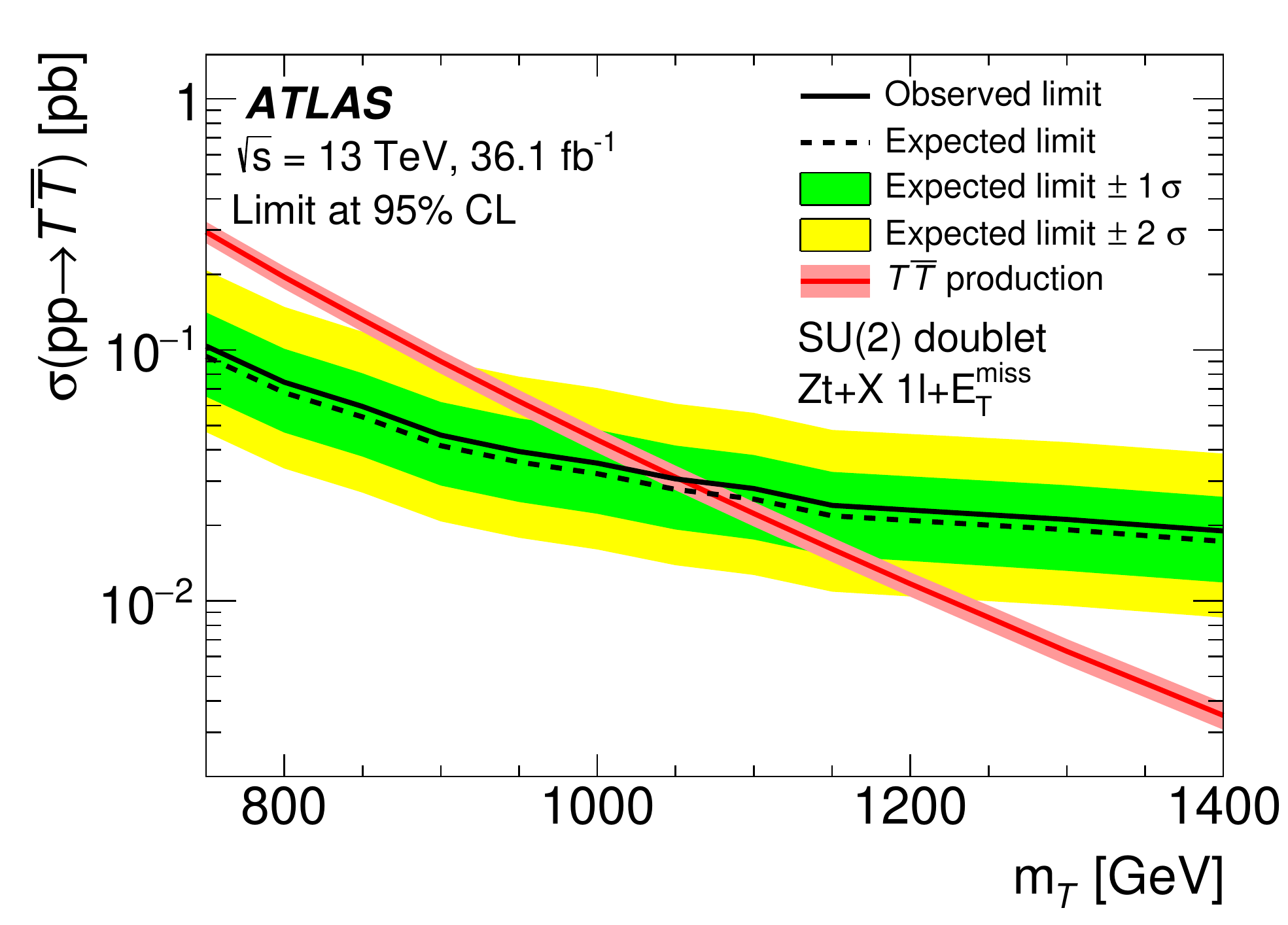}
	\caption{Observed and expected 95\% CL upper limit on the cross-section times branching ratio for VLT pair production as a function of the $T$ mass for $\BRto{Zt} = 100\%$ (top) and for branching ratios according to the singlet model~(bottom left) and the doublet model (bottom right). Contributions from the $X$ or $B$ quark in the $(X^{5/3}, T)$ or $(T, B)$ doublet models are neglected, leading to conservative limits. 
The thickness of the theory curve represents the theoretical uncertainty from PDFs, scale and the strong coupling constant \alphas.
}
	\label{fig:xs-limit}
\end{figure}

\begin{table}[tbp]
	\centering
	\caption{Observed and expected 95\% CL lower limits on the $T$ quark mass for the pure $T\rightarrow Zt$, the singlet model and the doublet model.
Contributions from the $X$ or $B$ quark in the $(X^{5/3}, T)$ or $(T, B)$ doublet models are neglected, leading to conservative limits.}
	\begin{tabular}{c|c|c}
		\toprule
		           Signal & Obs. 95\% CL  & Exp. 95\% CL  \\ 
		                      & lower mass limit &  lower mass limit \\ \midrule
		$T\rightarrow Zt$ &            1.16\,\TeV &            1.17\,\TeV \\
		          Singlet &            0.87\,\TeV &            0.89\,\TeV \\
		          Doublet &            1.05\,\TeV &            1.06\,\TeV \\ \bottomrule
	\end{tabular}
	\label{tab:mass-limit}
\end{table}

Figure~\ref{fig:mass-limit} shows the 95\% CL observed and expected lower limits on the signal mass as a function of the decay branching ratios. 
The highest sensitivity is found in the region of \mbox{$\BRto{Zt}\simeq100\%$}.
The sensitivity to the mixed $ZtHt$ decay mode is larger than to the $ZtWb$ decay mode. As an example, 
masses of $m_{T} < 1\,\TeV$ are excluded for $\BRto{Ht} < 65\%$ (70\% expected) in the former decay mode 
and for $\BRto{Wb} < 45\%$ (50\% expected) in the latter.

\begin{figure}[htb]
	\centering
	\includegraphics[width=.5\textwidth]{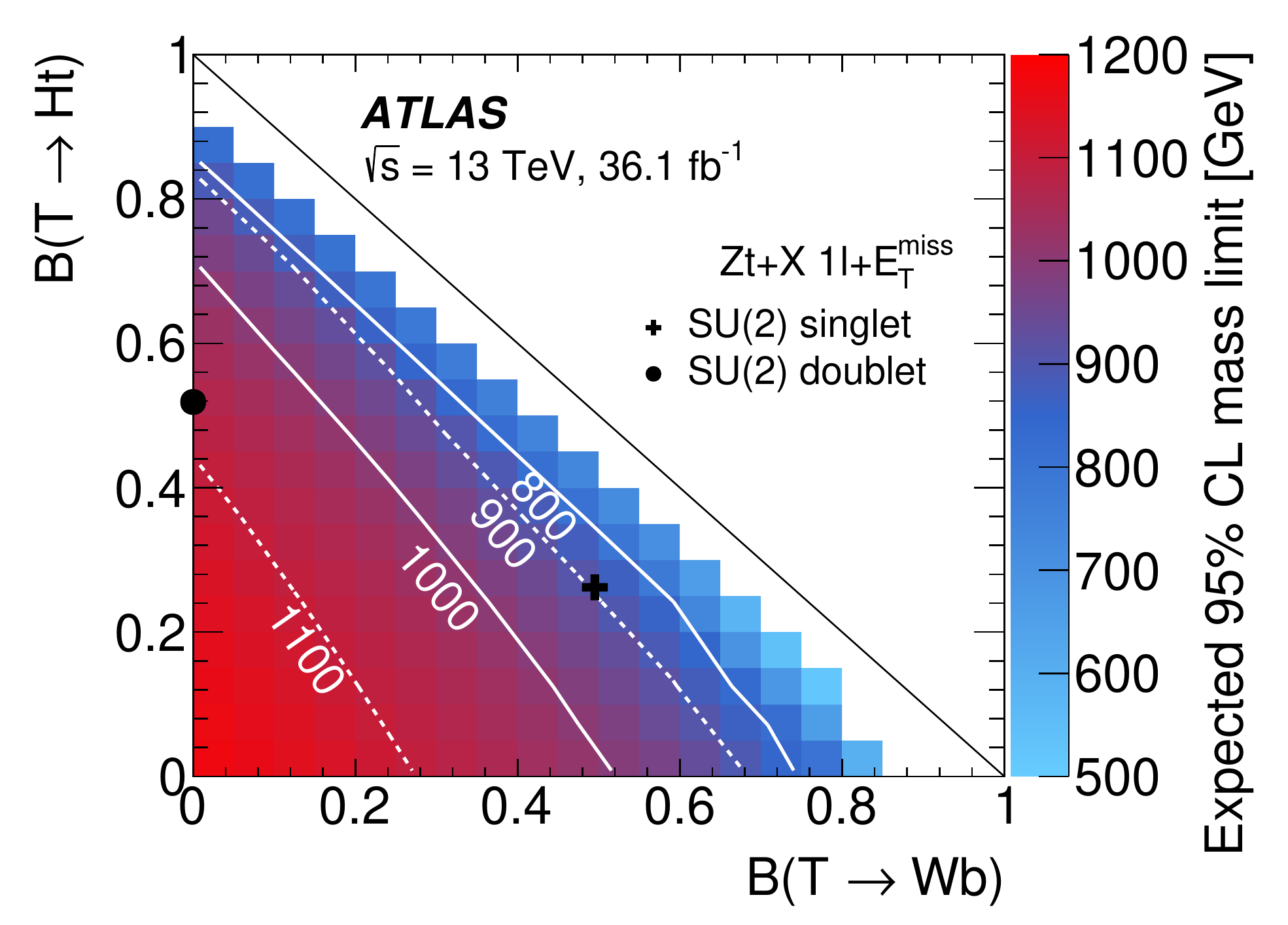}%
	\includegraphics[width=.5\textwidth]{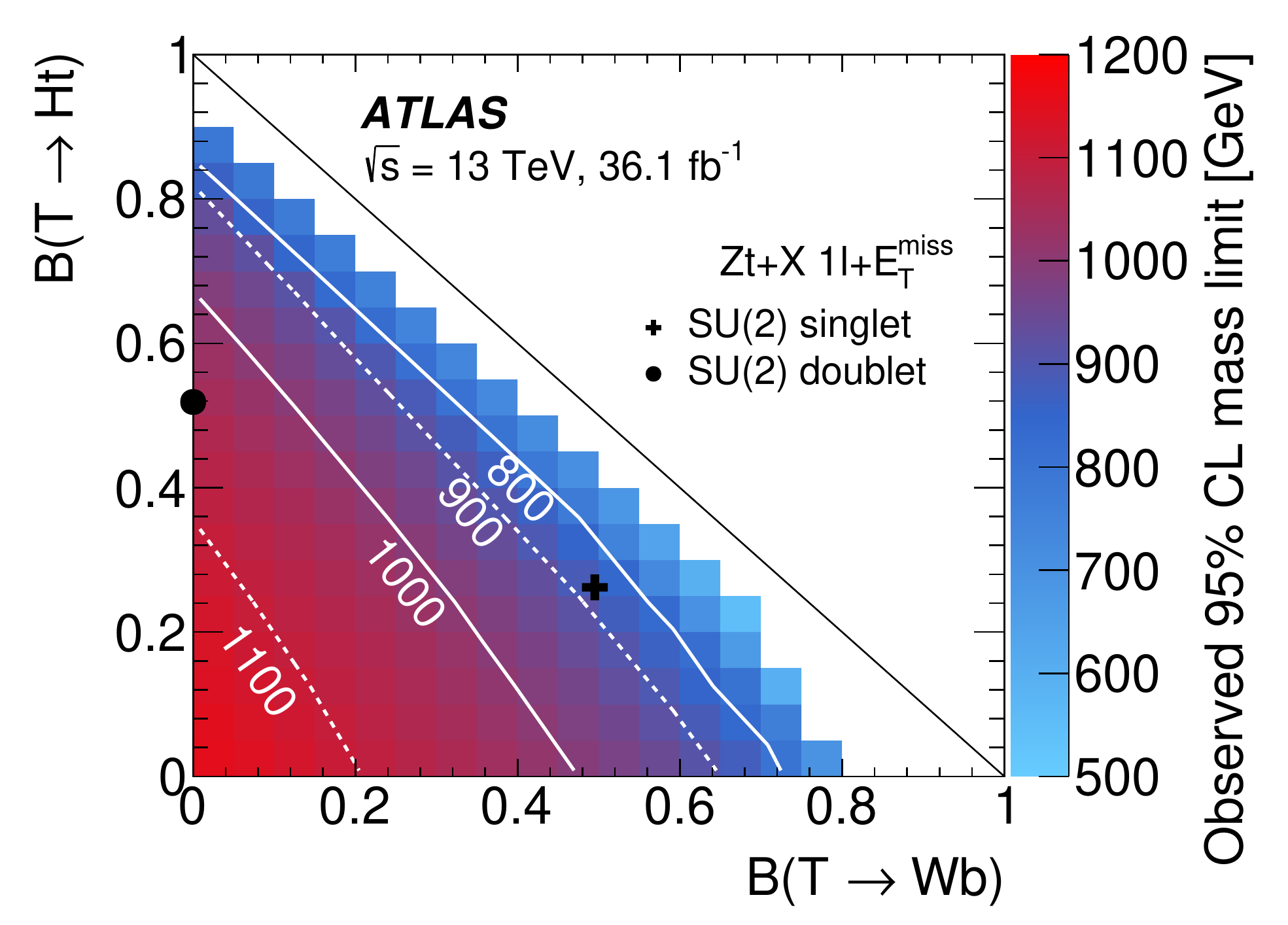}
	\caption{Expected (left) and observed (right) 95\% CL lower limit on the $T$ quark mass as a function of the decay branching ratios into $Wb$ and $Ht$. 
The markers indicate the branching ratios in the singlet and doublet models for masses above about 0.8\,\TeV, where they are approximately independent of the $T$ quark mass.}
	\label{fig:mass-limit}
\end{figure}


\FloatBarrier
\section{Conclusion}
\label{sec:conclusion}

A search for the pair production of vector-like top quarks in final states with exactly one lepton, at least four jets 
and high missing transverse momentum is presented. 
The analysis follows a cut and count strategy, with a single-bin signal region and dedicated control regions for the \ttbar and \Wjets backgrounds. 
The search is based on 36.1\,\ifb of $\sqrt{s}=13\,\TeV$ LHC $pp$ collision data recorded by ATLAS in 2015 and 2016. 
No significant excess above the SM background is observed in the signal region and 95\% CL limits are set on the vector-like top quark mass.
For the singlet and doublet models, masses below 870\,\GeV (890\,\GeV expected) and 1.05\,\TeV (1.06\,\TeV expected), respectively, are excluded. 
For the pure $Zt$ decay, lower limits are set on the
mass at 1.16\,\TeV (1.17\,\TeV expected). 
Limits are also set on the mass as a function of the decay branching ratios, excluding large parts of the parameter space for masses below 1\,\TeV.


\section*{Acknowledgements}


We thank CERN for the very successful operation of the LHC, as well as the
support staff from our institutions without whom ATLAS could not be
operated efficiently.

We acknowledge the support of ANPCyT, Argentina; YerPhI, Armenia; ARC, Australia; BMWFW and FWF, Austria; ANAS, Azerbaijan; SSTC, Belarus; CNPq and FAPESP, Brazil; NSERC, NRC and CFI, Canada; CERN; CONICYT, Chile; CAS, MOST and NSFC, China; COLCIENCIAS, Colombia; MSMT CR, MPO CR and VSC CR, Czech Republic; DNRF and DNSRC, Denmark; IN2P3-CNRS, CEA-DSM/IRFU, France; SRNSF, Georgia; BMBF, HGF, and MPG, Germany; GSRT, Greece; RGC, Hong Kong SAR, China; ISF, I-CORE and Benoziyo Center, Israel; INFN, Italy; MEXT and JSPS, Japan; CNRST, Morocco; NWO, Netherlands; RCN, Norway; MNiSW and NCN, Poland; FCT, Portugal; MNE/IFA, Romania; MES of Russia and NRC KI, Russian Federation; JINR; MESTD, Serbia; MSSR, Slovakia; ARRS and MIZ\v{S}, Slovenia; DST/NRF, South Africa; MINECO, Spain; SRC and Wallenberg Foundation, Sweden; SERI, SNSF and Cantons of Bern and Geneva, Switzerland; MOST, Taiwan; TAEK, Turkey; STFC, United Kingdom; DOE and NSF, United States of America. In addition, individual groups and members have received support from BCKDF, the Canada Council, CANARIE, CRC, Compute Canada, FQRNT, and the Ontario Innovation Trust, Canada; EPLANET, ERC, ERDF, FP7, Horizon 2020 and Marie Sk{\l}odowska-Curie Actions, European Union; Investissements d'Avenir Labex and Idex, ANR, R{\'e}gion Auvergne and Fondation Partager le Savoir, France; DFG and AvH Foundation, Germany; Herakleitos, Thales and Aristeia programmes co-financed by EU-ESF and the Greek NSRF; BSF, GIF and Minerva, Israel; BRF, Norway; CERCA Programme Generalitat de Catalunya, Generalitat Valenciana, Spain; the Royal Society and Leverhulme Trust, United Kingdom.

The crucial computing support from all WLCG partners is acknowledged gratefully, in particular from CERN, the ATLAS Tier-1 facilities at TRIUMF (Canada), NDGF (Denmark, Norway, Sweden), CC-IN2P3 (France), KIT/GridKA (Germany), INFN-CNAF (Italy), NL-T1 (Netherlands), PIC (Spain), ASGC (Taiwan), RAL (UK) and BNL (USA), the Tier-2 facilities worldwide and large non-WLCG resource providers. Major contributors of computing resources are listed in Ref.~\cite{ATL-GEN-PUB-2016-002}.

\clearpage

\FloatBarrier

\printbibliography

\ifdefined\IncludeAuxMat
\clearpage
\appendix
\part*{Auxiliary material}
\addcontentsline{toc}{part}{Auxiliary material}

\input{texfiles/auxmat}
\fi

\FloatBarrier
\newpage
\begin{flushleft}
{\Large The ATLAS Collaboration}

\bigskip

M.~Aaboud$^\textrm{\scriptsize 137d}$,
G.~Aad$^\textrm{\scriptsize 88}$,
B.~Abbott$^\textrm{\scriptsize 115}$,
O.~Abdinov$^\textrm{\scriptsize 12}$$^{,*}$,
B.~Abeloos$^\textrm{\scriptsize 119}$,
S.H.~Abidi$^\textrm{\scriptsize 161}$,
O.S.~AbouZeid$^\textrm{\scriptsize 139}$,
N.L.~Abraham$^\textrm{\scriptsize 151}$,
H.~Abramowicz$^\textrm{\scriptsize 155}$,
H.~Abreu$^\textrm{\scriptsize 154}$
% ...
\bigskip
\\
% ...
$^{12}$ Institute of Physics, Azerbaijan Academy of Sciences, Baku, Azerbaijan\\
% ...
$^{88}$ CPPM, Aix-Marseille Universit{\'e} and CNRS/IN2P3, Marseille, France\\
% ...
$^{115}$ Homer L. Dodge Department of Physics and Astronomy, University of Oklahoma, Norman OK, United States of America\\
% ...
$^{119}$ LAL, Univ. Paris-Sud, CNRS/IN2P3, Universit{\'e} Paris-Saclay, Orsay, France\\
% ...
$^{137}$ $^{(a)}$ Facult{\'e} des Sciences Ain Chock, R{\'e}seau Universitaire de Physique des Hautes Energies - Universit{\'e} Hassan II, Casablanca; $^{(b)}$ Centre National de l'Energie des Sciences Techniques Nucleaires, Rabat; $^{(c)}$ Facult{\'e} des Sciences Semlalia, Universit{\'e} Cadi Ayyad, LPHEA-Marrakech; $^{(d)}$ Facult{\'e} des Sciences, Universit{\'e} Mohamed Premier and LPTPM, Oujda; $^{(e)}$ Facult{\'e} des sciences, Universit{\'e} Mohammed V, Rabat, Morocco\\
% ...
$^{139}$ Santa Cruz Institute for Particle Physics, University of California Santa Cruz, Santa Cruz CA, United States of America\\
% ...
$^{151}$ Department of Physics and Astronomy, University of Sussex, Brighton, United Kingdom\\
% ...
$^{154}$ Department of Physics, Technion: Israel Institute of Technology, Haifa, Israel\\
$^{155}$ Raymond and Beverly Sackler School of Physics and Astronomy, Tel Aviv University, Tel Aviv, Israel\\
% ...
$^{161}$ Department of Physics, University of Toronto, Toronto ON, Canada\\
% ...
$^{a}$ Also at Department of Physics, King's College London, London, United Kingdom\\
$^{b}$ Also at Institute of Physics, Azerbaijan Academy of Sciences, Baku, Azerbaijan\\
$^{c}$ Also at Novosibirsk State University, Novosibirsk, Russia\\
$^{d}$ Also at TRIUMF, Vancouver BC, Canada\\
$^{e}$ Also at Department of Physics {\&} Astronomy, University of Louisville, Louisville, KY, United States of America\\
% ...
$^{*}$ Deceased
\end{flushleft}


\end{document}